\documentstyle[12pt,fleqn,rotate,psfig]{article}
\newcommand{\be}{\begin{equation}}
\newcommand{\ee}{\end{equation}}
\newcommand{\bea}{\begin{eqnarray}}
\newcommand{\eea}{\end{eqnarray}}

\newcommand{\Psb}{\bar{\Psi}}
\newcommand{\psb}{\bar{\psi}}
\begin{document}
\title{Correlation Dynamics of\\
Yukawa-theory in 1+1 dimensions\footnote{supported by DFG, GSI Darmstadt}}
\author{S. Juchem, W. Cassing and J.M. H\"auser\\
Institut f\"ur Theoretische Physik, Universit\"at Gie\ss en\\
35392 Gie\ss en, Germany}
\maketitle
\begin{abstract}
Using the method of correlation dynamics we investigate the 
properties of a field-theory for fermions and 
scalar bosons coupled via a Yukawa interaction. 
Within this approach, which consists in an expansion of full 
equal-time Green functions into connected equal-time 
Green functions and a corresponding truncation of the 
hierarchy of equations of motion we carry out calculations up 
to 4th order in the connected Green functions and evaluate 
the effective potential of the  theory in 1+1 dimensions on a 
torus.
Comparing the different approximations we find a 
strong influence of the connected 4-point functions on the 
properties of the system.
\end{abstract}
\bigskip
\newpage
\section{Introduction}
\label{introduction}
Up to now the problem of strongly interacting fermion and 
boson field-theories is not solved in a satisfying manner 
and there is a need for nonperturbative approximation schemes.
In the last decades many non-perturbative approaches were 
developed such as variational methods \cite{st85,fns86},
Dyson-Schwinger-calculations \cite{kk85,ba96} or the coupled 
cluster expansion \cite{arp83,fkk87}.
Among the approaches that are based on techniques known
from many-particle physics is the method of correlation dynamics, 
which was successfully applied to the nonrelativistic nuclear 
many-particle problem \cite{bct92,cpp93,gkrt92} and for the
description of interacting boson systems \cite{hcp95a}. 
Within a formulation in terms of connected equal-time 
Green functions the problem of ground-state symmetry
breaking was studied in detail for $\Phi^4$-theory
in $1+1$ and $2+1$ space-time dimensions \cite{hcp95b,pht95}
as well as the convergence properties of the 
truncation scheme applied \cite{phc97}.
In this article the correlation dynamical approach is  
extended from a pure boson theory to a field theory containing 
both bosonic and fermionic degrees of freedom, i.e. 
Yukawa-theory in $1+1$ space-time dimensions.

The paper is organized as follows: 
In Section 2 we present the connected Green function approach
for Yukawa-theory in $1+1$ space-time dimensions and
discuss the correlation dynamical hierarchy for connected 
equal-time Green functions in different truncation schemes.
Furtheron, we study the divergence structure
of the Yukawa-theory in $1+1$ dimensions and renormalize
as in a comparable Gaussian-Effective-Potential-(GEP)-approach
\cite{shr86}.
In Section 3 we show the results obtained in our numerical
calculations for different truncation schemes
up to the level of connected 4-point functions.
A summary is given in Section 4 while the Appendices contain
the higher order cluster expansions and all correlation dynamical 
equations of motion up to fourth order.          
\section{Correlation dynamics of Yukawa-theory}
\label{theory}
\subsection{The equations of motion for connected 
equal-time \\ Green functions}
In the present work we investigate the Yukawa-theory in 
$1+1$ space-time dimensions given by the Lagrangian,
\bea
{\cal L} & = & \frac{1}{2} \partial_\mu \Phi \ \partial^\mu \Phi
-\frac{1}{2} m_B^2 \Phi^2
+\Psb (i{\gamma^\mu}{\partial_\mu} -M_B) \Psi
-g_B \Psb \Psi \Phi \; , 
\label{lagrangian}
\eea
or the Hamiltonian density, 
\bea
{\cal H} & = & \frac{1}{2} \Pi^2
+\frac{1}{2} (\vec {\nabla} \Phi )^2
+\frac{1}{2} m_B^2 \Phi^2
+\Psb (-i{\vec {\gamma}}{\vec {\nabla}} +M_B) \Psi
+g_B \Psb {\Psi} \Phi \; . 
\label{hamiltonian}
\eea
Here $m_B$ and $M_B$ denote the bare masses of the boson 
and the fermion fields, respectively, while $g_B$ represents 
the bare Yukawa-coupling. 
The fields themselves have to be considered as bare fields 
where we have dropped the corresponding indices.
They fulfill equal-time commutation relations 
for the boson fields
\be
\lbrack \Phi({\vec {x}},t), \Pi({\vec {y}},t) \rbrack
\ = \
i \ {\delta}^{(\nu)} (\vec {x} - \vec {y}) \; ,
\ee
\be
\lbrack \Phi({\vec {x}},t), \Phi({\vec {y}},t) \rbrack
\ = \
\lbrack \Pi({\vec {x}},t), \Pi({\vec {y}},t) \rbrack \ = \ 0 
\ee
and equal-time anticommutation relations for the
fermion fields
\be
\lbrace \Psb_a ({\vec {x}},t), \Psi_b ({\vec {y}},t) \rbrace
\ = \
{\gamma}_{ab}^0 \ {\delta}^{(\nu)} (\vec {x} - \vec {y}) \; ,
\ee
\be
\lbrace \Psb_a ({\vec {x}},t), \Psb_b ({\vec {y}},t) \rbrace
\ = \
\lbrace \Psi_a ({\vec {x}},t), \Psi_b ({\vec {y}},t) \rbrace \ = \ 0 \; ,
\ee
where $a$ and $b$ denote the components of the spinor.
Making use of these relations one obtains the time evolution 
of any operator $\hat{O}$ via the Heisenberg equation
\bea
i {\partial}_t \ \hat{O} & = & \lbrack \, \hat{O} , H \, \rbrack \; .
\label{heisenbergeom}
\eea
The equations of motion for the field operators read:
\bea
\partial_t \ \Phi (\vec{x})
& = & \Pi (\vec{x}) \; ,
\label{uf}
\eea
\bea
\partial_t \ \Pi (\vec{x})
& = & ({\vec {\nabla}}^2_x - m^2_B) \, \Phi (\vec{x})
    - g_B \Psb_l (\vec{x}) \Psi_l (\vec{x}) \; ,
\label{up}
\eea
\bea
\partial_t \ \Psb_a (\vec{x})
& = & \phantom{- \,} {\vec {\alpha}}_{la} {\vec {\nabla}}_x \Psb_l (\vec{x})
    + i M_B \beta_{la} \Psb_l (\vec{x})
    + i g_B \beta_{la} \Psb_l (\vec{x}) \Phi (\vec{x}) \; ,
\label{uw}
\eea
\bea
\partial_t \ \Psi_b (\vec{x})
& = & - \, {\vec {\alpha}}_{bl} {\vec {\nabla}}_x \Psi_l (\vec{x})
      - i M_B \beta_{bl} \Psi_l (\vec{x})
      - i g_B \beta_{bl} \Psi_l (\vec{x}) \Phi (\vec{x}) \; ,
      \nonumber
\label{uv}
\eea
where all operators are considered at the same time $t$ which
is suppressed in our notation. 
In 1+1 dimensions the Dirac-matrices are given by the repesentation
\bea
\beta \; = \; \gamma^0 \; = \;
\left(
\begin{array}{rr}
1 & 0 \\
0 & -1 
\end{array}
\right) \; ,
\qquad
\alpha^1 \; = \;
\left(
\begin{array}{rr}
0 & 1 \\
1 & 0 
\end{array}
\right) \; ,
\qquad
\gamma^1 \; = \; 
\beta \, \alpha^1 \; = \;
\left(
\begin{array}{rr}
0 & 1 \\
-1 & 0 
\end{array}
\right) \; .
\eea
For any arbitrary product of these field operators the time
evolution is obtained correspondingly.
The boson field momentum $\Pi$ is considered explicitly in 
order to end up with equations of motion of first order in time.
Due to the Yukawa-coupling term the time evolution
of any field operator product, which consists at least of two 
fermion fields or one boson field momentum, is influenced
by field operator products of the next higher order. 

By taking the expectation value of these operator identities 
we end up with an infinite hierarchy of equations of motion for 
equal-time Green functions which is equivalent to the 
BBGKY\footnote{Bogoliubov-Born-Green-Kirkwood-Yvon}-density-matrix-hierarchy 
of ordinary many-body physics \cite{bal75}.
For the lowest orders we find:
\begin{eqnarray}
\partial_t \; \langle \, \Phi (\vec{x}_1) \Phi (\vec{x}_2) \, \rangle
& = & \langle \, \Pi (\vec{x}_1) \Phi (\vec{x}_2) \, \rangle
    + \langle \, \Phi (\vec{x}_1) \Pi (\vec{x}_2) \, \rangle \; ,
\label{uff}
\end{eqnarray}
\begin{eqnarray}
\partial_t \; \langle \, \Pi (\vec{x}_1) \Phi (\vec{x}_2) \, \rangle
& = & ({\vec {\nabla}}^2_{x_1} - m^2_B) \;
      \langle \, \Phi (\vec{x}_1) \Phi (\vec{x}_2) \, \rangle
    + \langle \, \Pi (\vec{x}_1) \Pi (\vec{x}_2) \, \rangle
    \\
& & - \ g_B \; \langle \, \Psb_l (\vec{x}_1) \Psi_l (\vec{x}_1) 
                       \Phi(\vec{x}_2) \, \rangle
    \nonumber  \; ,
\label{upf}
\end{eqnarray}
\begin{eqnarray}
\partial_t \; \langle \, \Pi (\vec{x}_1) \Pi (\vec{x}_2) \, \rangle
& = & ({\vec {\nabla}}^2_{x_1} - m^2_B) \; 
      \langle \, \Phi (\vec{x}_1) \Pi (\vec{x}_2) \, \rangle
    + ({\vec {\nabla}}^2_{x_2} - m^2_B) \;
      \langle \, \Pi (\vec{x}_1) \Phi (\vec{x}_2) \, \rangle
    \\
& & - \ g_B \; \langle \, \Psb_l (\vec{x}_1) \Psi_l (\vec{x}_1) 
                       \Pi(\vec{x}_2) \, \rangle
    - g_B \;  \langle \, \Psb_l (\vec{x}_2) \Psi_l (\vec{x}_2) 
                       \Pi(\vec{x}_1) \, \rangle
    \nonumber \; ,
\label{upp}
\end{eqnarray}
\begin{eqnarray}
\partial_t \; \langle \, \Psb_a (\vec{x}_1) \Psi_b (\vec{x}_2) \, \rangle
& = & {\vec {\alpha}}_{la}
      {\vec {\nabla}}_{x_1} \, 
      \langle \, \Psb_l (\vec{x}_1) \Psi_b (\vec{x}_2) \, \rangle
    - {\vec {\alpha}}_{bl}
      {\vec {\nabla}}_{x_2} \,
      \langle \, \Psb_a (\vec{x}_1) \Psi_l (\vec{x}_2) \, \rangle
      \\
& & + \ i M_B \beta_{la} \; 
      \langle \, \Psb_l (\vec{x}_1) \Psi_b (\vec{x}_2) \, \rangle
    - i M_B \beta_{bl} \;
      \langle \, \Psb_a (\vec{x}_1) \Psi_l (\vec{x}_2) \, \rangle
      \nonumber\\
& & + \ i g_B \beta_{la} \; 
      \langle \, \Psb_l (\vec{x}_1) \Psi_b (\vec{x}_2) 
                 \Phi (\vec{x}_1) \, \rangle
      \nonumber\\
& &  - \ i g_B \beta_{bl} \; 
      \langle \, \Psb_a (\vec{x}_1) \Psi_l (\vec{x}_2) 
                 \Phi (\vec{x}_2) \, \rangle
    \nonumber \; .
\label{uwv}
\end{eqnarray}
In order to solve this infinite set of equations a 
truncation scheme must be applied.
The ansatz of correlation dynamics consists of two steps:
i) all full equal-time Green functions are replaced by sums 
of products of connected equal-time Green functions via a 
cluster expansion, which leads to a -- still infinite -- 
hierarchy for the connected Green functions, ii) a truncation 
scheme is applied to this hierarchy by neglecting all connected 
equal-time Green functions of order $n>N$.
This closed set of equations of motion for 
connected equal-time Green functions is denoted 
as the {\it correlation dynamical $N$-point approximation}.

Before deriving this set for the Yukawa-theory we separate 
the classical part from the boson field, since here we are 
primarily interested in the effective potential of the theory.
The boson field is written as
\be
\Phi = \Phi_0 + \hat{\Phi} \; ,
\label{sep1}
\ee
where $\Phi_0$ is the classical contribution given by
$\langle \Phi \rangle = \Phi_0$ and $\hat{\Phi}$ the 
pure quantum mechanical part with vanishing 
expectation value $\langle \hat{\Phi} \rangle = 0$.
With this separation the effective potential of the field theory 
is equivalent to the minimal energy density in the subspace 
of fixed vacuum expectation value $\Phi_0$, i.e.  
\begin{eqnarray}
V_{{\rm eff}}(\Phi_0)
 = {{\rm min}\atop\{\Xi\}}
\langle \Xi | {\cal{H}} | \Xi \rangle \; , 
\quad 
\quad {\rm with}
\quad
\langle \Xi|\Phi|\Xi \rangle =\Phi_0 \; ,
\quad
\langle \Xi|\Xi \rangle =1 \; .
\label{veffvar}
\end{eqnarray}
Using (\ref{sep1}) the modified Lagrangian reads
\be
{\cal L} = \frac{1}{2}
\partial_\mu \hat{\Phi} \ \partial^\mu \hat{\Phi}
-\frac{1}{2} m_B^2 (\hat{\Phi} + \Phi_0)^2
+\Psb (i{\gamma^\mu}{\partial_\mu} -M_B -g_B \Phi_0) \Psi
-g_B \Psb {\Psi} \hat{\Phi} \; ,
\label{lagrangiansep}
\ee
and the corresponding Hamiltonian density is given by
\be
{\cal H} = \frac{1}{2} \hat{\Pi}^2
+\frac{1}{2} (\vec {\nabla} \hat{\Phi} )^2
+\frac{1}{2} m_B^2 (\hat{\Phi} + \Phi_0)^2
+\Psb (-i{\vec {\gamma}}{\vec {\nabla}} +M_B +g_B \Phi_0) \Psi
+g_B \Psb {\Psi} \hat{\Phi} \; .
\label{hamiltoniansep}
\ee
Since we take $\Phi_0$ to be fixed, i.e. $\partial_t \Phi_0 = 0$,
we have $\Pi = \partial_t \Phi = \partial_t \hat{\Phi} = 
\hat{\Pi}$.
As a consequence the Hamiltonian density now contains a 
contribution which is generated by the interaction part and 
has the same structure as the fermion mass term.
The modified equations of motion for the full equal-time
Green functions are obtained by replacing $M_B$  
by $M_B + g_B \Phi_0$ and the boson field $\Phi$ by its 
quantum part $\hat{\Phi}$.

The cluster expansion is derived from the relation
between the generating functionals of full and connected Green 
functions and by taking the well defined equal-time limit.
For the bosonic case this is presented in detail
in Ref. \cite{hcp95b} in context of $\Phi^4$-theory and thus will not 
be repeated here. 
We only show the expansion for the lowest order 
boson Green functions (with field operators 
$\hat{O}_i(\vec{x}_i) \in \{ \Phi(\vec{x}_i) , \Pi(\vec{x}_i) \}$ ):
\bea
\langle \, \hat{O}_1(\vec{x}_1) \, \rangle \; = \; 
\langle \, \hat{O}_1(\vec{x}_1) \, \rangle_c \; ,
\eea
\bea
\langle \, \hat{O}_1(\vec{x}_1) \, \hat{O}_2(\vec{x}_2) \, \rangle 
\; = \;
\langle \, \hat{O}_1(\vec{x}_1) \, \hat{O}_2(\vec{x}_2) \, \rangle_c 
\; + \;
\langle \, \hat{O}_1(\vec{x}_1) \, \rangle_c \; 
\langle \, \hat{O}_2(\vec{x}_2) \, \rangle_c \; ,
\label{tffclust}
\eea
\bea
\lefteqn{
\langle \, \hat{O}_1(\vec{x}_1) \, \hat{O}_2(\vec{x}_2) \, 
           \hat{O}_3(\vec{x}_3) \, \rangle \; = \;
\langle \, \hat{O}_1(\vec{x}_1) \, \hat{O}_2(\vec{x}_2) \, 
           \hat{O}_3(\vec{x}_3) \, \rangle_c }
\nonumber\\
& &+ \; \langle \, \hat{O}_1(\vec{x}_1) \, 
                   \hat{O}_2(\vec{x}_2) \, \rangle_c \; 
        \langle \, \hat{O}_3(\vec{x}_3) \, \rangle_c \;
   + \; \langle \, \hat{O}_1(\vec{x}_1) \, 
                   \hat{O}_3(\vec{x}_3) \, \rangle_c \;
        \langle \, \hat{O}_2(\vec{x}_2) \, \rangle_c
\nonumber\\
& &+ \; \langle \, \hat{O}_2(\vec{x}_2) \, 
                   \hat{O}_3(\vec{x}_3) \, \rangle_c \; 
        \langle \, \hat{O}_1(\vec{x}_1) \, \rangle_c \;
   + \; \langle \, \hat{O}_1(\vec{x}_1) \, \rangle_c \; 
        \langle \, \hat{O}_2(\vec{x}_2) \, \rangle_c \;
        \langle \, \hat{O}_3(\vec{x}_3) \, \rangle_c \; .
\label{tfffclust}
\eea
In the fermionic sector the source terms appearing in the 
generating functionals are anticommuting Grassmann fields. 
Thus the cluster expansions differ from the boson case and 
read in the lowest orders:
\bea
\langle \, \Psb(\vec{x}_1) \, \Psi(\vec{x}_2) \, \rangle \; = \;
\langle \, \Psb(\vec{x}_1) \, \Psi(\vec{x}_2) \, \rangle_c \; ,
\label{twvclust}
\eea
\bea
\lefteqn{
\langle \, \Psb(\vec{x}_1) \, \Psb(\vec{x}_2) \,
           \Psi(\vec{x}_3) \, \Psi(\vec{x}_4) \, \rangle \; = \;
\langle \, \Psb(\vec{x}_1) \, \Psb(\vec{x}_2) \, 
           \Psi(\vec{x}_3) \, \Psi(\vec{x}_4) \, \rangle_c }
\nonumber\\
& & + \; \langle \, \Psb(\vec{x}_1) \, \Psi(\vec{x}_4) \, \rangle_c 
      \; \langle \, \Psb(\vec{x}_2) \, \Psi(\vec{x}_3) \, \rangle_c
\nonumber\\
& & - \; \langle \, \Psb(\vec{x}_1) \, \Psi(\vec{x}_3) \, \rangle_c 
      \; \langle \, \Psb(\vec{x}_2) \, \Psi(\vec{x}_4) \, \rangle_c \; .
\label{twwvvclust}
\eea
Due to the absence of fermionic n-point functions of odd order
the full and the connected 2-point function are equivalent.
Therefore the index $\cdot_c$ for the connected Green
functions can be dropped for $\langle \Psb \Psi \rangle$ 
as in case of the boson 1-point function.
For the mixed 3-point functions we find:
\bea
\langle \, \Psb(\vec{x}_1) \, \Psi(\vec{x}_2) \, 
           \hat{O}_3(\vec{x}_3)\} \, \rangle
& = & 
\langle \, \Psb(\vec{x}_1) \, \Psi(\vec{x}_2) \,
           \hat{O}_3(\vec{x}_3)\} \, \rangle_c
      \nonumber\\
& & + \; \langle \, \Psb(\vec{x}_1) \, \Psi(\vec{x}_2) \, \rangle_c \;
         \langle \, \hat{O}_3(\vec{x}_3) \, \rangle \; ;
\label{twvfclust}
\eea
the higher orders up to the 5-point-level are given
in Appendix A.

An insertion of the cluster expansions into the hierarchy of 
equations of motion for full equal-time Green functions 
leads to a corresponding hierarchy for connected 
equal-time Green functions.
In the present work three different truncation prescriptions
are applied to this coupled set of equations,
i.e. including all connected equal-time Green functions up to 
the 2-point-, the 3-point- and the 4-point-level.
These approximations are denoted as $Y_{1+1}CD(2)$-, 
$Y_{1+1}CD(3)$- and $Y_{1+1}CD(4)$-approximation.
As examples for the equations of motion for connected equal-time
Green functions we present those of lowest order:
\bea
\partial_t \, \langle \, \hat{\Phi} (\vec{x}_1) \, 
                         \hat{\Phi} (\vec{x}_2) \, \rangle_c
& = &       \langle \, \hat{\Pi} (\vec{x}_1)  \, 
                         \hat{\Phi} (\vec{x}_2) \, \rangle_c 
\; + \;       \langle \, \hat{\Phi} (\vec{x}_1) \, 
                         \hat{\Pi} (\vec{x}_2)  \, \rangle_c \; ,
\label{cff}
\eea
\bea
\partial_t \, \langle \, \hat{\Pi} (\vec{x}_1)  \, 
                         \hat{\Phi} (\vec{x}_2) \, \rangle_c
& = & \phantom{+ \ } t(\vec{x}_1) \;
      \langle \, \hat{\Phi} (\vec{x}_1) \, 
                 \hat{\Phi} (\vec{x}_2) \, \rangle_c 
\; + \;  
      \langle \, \hat{\Pi} (\vec{x}_1)  \, 
                 \hat{\Pi} (\vec{x}_2)  \, \rangle_c
      \\
& & - \; g_B \: \langle \, \Psb_l (\vec{x}_1) \, \Psi_l (\vec{x}_1) \,
      \hat{\Phi}(\vec{x}_2) \, \rangle_c \; ,
      \nonumber
\label{cpf}
\eea
\bea
\partial_t \, \langle \, \hat{\Pi} (\vec{x}_1) \, 
                         \hat{\Pi} (\vec{x}_2) \, \rangle_c
& = & \phantom{+ \ } t(\vec{x}_1) \;
      \langle \, \hat{\Phi} (\vec{x}_1) \, 
                 \hat{\Pi} (\vec{x}_2)  \, \rangle_c
\; + \;              t(\vec{x}_2) \;
      \langle \, \hat{\Pi} (\vec{x}_1)  \, 
                 \hat{\Phi} (\vec{x}_2) \, \rangle_c
      \\
& & - \;  g_B \: 
      \langle \, \Psb (\vec{x}_1) \, \Psi (\vec{x}_1) \,
             \hat{\Pi}(\vec{x}_2) \, \rangle_c
 \; - \;  g_B \: 
      \langle \, \Psb (\vec{x}_2) \, \Psi (\vec{x}_2) \, 
             \hat{\Pi}(\vec{x}_1) \, \rangle_c \; ,
      \nonumber
\label{cpp}
\eea
\bea
\partial_t \, \langle \, \Psb_a(\vec{x}_1) \, 
                         \Psi_b(\vec{x}_2) \, \rangle_c
& = & \phantom{+ \ } \tilde{t}(\vec{x}_1)_{la} \;
      \langle \, \Psb_l(\vec{x}_1) \, \Psi_b(\vec{x}_2) \, \rangle_c
\; - \;              \tilde{t}(\vec{x}_2)_{lb} \;
      \langle \, \Psb_a(\vec{x}_1) \, \Psi_l(\vec{x}_2) \, \rangle_c 
      \\
& & + \;  i g_B \beta_{la} \;
      \langle \, \Psb_l(\vec{x}_1) \, \Psi_b(\vec{x}_2) \, 
             \hat{\Phi}(\vec{x}_1) \, \rangle_c
      \nonumber\\
& & - \;  i g_B \beta_{bl} \;
      \langle \, \Psb_a(\vec{x}_1) \, \Psi_l(\vec{x}_2) \,
             \hat{\Phi}(\vec{x}_2) \, \rangle_c \; ,
      \nonumber
\label{cwv}
\eea
where the connected 3-point functions are neglected 
in the limit $Y_{1+1}CD(2)$.
In (27)-(29)
the following operators have been introduced for the 
bosonic part
\be
t(\vec{x}) \: := \: ({\vec {\nabla}}^2_{x} - m^2_B)
\label{tbos}
\ee
and the fermionic part
\be
\tilde{t}(\vec{x})_{ab} 
\: := \:  {\vec {\alpha}}_{ab} {\vec {\nabla}}_{x}
\: +  \: i ( M_B + g_B \Phi_0 ) \beta_{ab} \; .
\label{tfer}
\ee
With regard to a numerical treatment we expand all 
field operators in a basis of orthonormal single-particle 
wavefunctions.
Here we use a set of plane waves in a box of size $L$ 
with periodic boundary conditions.
This choice allows for an explicit exploitation of the 
translational invariance of the theory. 
The expansion of the field operators is given by:
\bea
\hat{\Phi}({\vec x})   & = & \sum_\alpha \ \varphi_\alpha \; 
\chi^{ }_\alpha ({\vec x}) \; ,
 \nonumber\\
\hat{\Pi}({\vec x})    & = & \sum_\alpha \ \pi_\alpha \; 
\chi^{ }_\alpha ({\vec x}) \; ,
 \nonumber\\
\Psb_a({\vec x}) & = & \sum_\alpha \ \psb_{a\alpha} \; 
\chi^{ }_\alpha ({\vec x}) \; ,
 \nonumber\\
\Psi_b({\vec x}) & = & \sum_\alpha \ \psi_{b\alpha} \; 
\chi^{ }_\alpha ({\vec x}) \; ,
\label{opbexp}
\eea
where the elements of the single-particle basis $\chi^{ }_\alpha$
obey the following orthonormality relations in $\nu + 1$ 
space-time dimensions:
\bea
\int d^{\nu}\!x \,\ \chi^*_{\alpha}({\vec x}) \ \chi^{ }_{\beta}({\vec x})
\; =\; \delta_{\alpha\beta} \; .
\label{opbortho}
\eea
The greek indices refer to the basis elements and the latin 
indices represent the spinor components in (\ref{opbexp}).
One straightforwardly obtains the equations of motion for 
the expansion coefficients of the connected equal-time 
Green functions. 
Introducing the abbreviations 
\be
\langle \alpha | \lambda_1 \lambda_2 \rangle \; = \;
\int d^{\nu}\!x \; \chi^*_\alpha ({\vec x})
\ \chi^{ }_{\lambda_1}({\vec x}) \ \chi^{ }_{\lambda_2}({\vec x}) \; ,
\label{opbcontr}
\ee
\bea
t_{\alpha \beta} & = &
\int d^{\nu}\!x \; \chi^*_\alpha ({\vec x}) \:
\left[ \, {\vec \nabla}^2_{\!x} -m_B^2 \,
\right]
\:
\chi^{ }_\beta (\vec{x})
\nonumber\\
& & = \int d^{\nu}\!x \; \chi^*_\alpha ({\vec x}) \;
t(\vec{x}) \;
\chi^{ }_\beta (\vec{x}) \; ,
\label{opbkinboson}
\eea
\bea
\tilde{t}_{\alpha \beta}^{a b} & = &
\int d^{\nu}\!x \; \chi^*_\alpha ({\vec x}) \:
\left[ \, \vec{\alpha}_{ab} {\vec \nabla}\!_x
       + i (M_B + g_B \Phi_0) \ \beta_{ab} \,
\right] \:
\chi^{ }_\beta (\vec{x})
\nonumber\\
& & = \int d^{\nu}\!x \; \chi^*_\alpha ({\vec x}) \;
\tilde{t}_{ab}(\vec{x}) \;
\chi^{ }_\beta (\vec{x}) \; ,
\label{opbkinfermion}
\eea
they read in lowest order:
\begin{eqnarray}
\partial_t \, \langle \, \varphi_\alpha \, \varphi_\beta \, \rangle_c
& = & \langle \, \pi_\alpha \, \varphi_\beta \, \rangle_c
\: + \:  \langle \, \varphi_\alpha \, \pi_\beta \, \rangle_c \; ,
\label{opbff}
\end{eqnarray}
\begin{eqnarray}
\partial_t \, \langle \, \pi_\alpha \, \varphi_\beta \, \rangle_c
& = & \langle \, \pi_\alpha \, \pi_\beta \, \rangle_c
\: + \:  \sum_{\lambda} \ t_{\alpha \lambda} \
      \langle \, \varphi_\lambda \, \varphi_\beta \, \rangle_c
      \nonumber\\
& & - \: g_B \sum_{\lambda_1 \lambda_2} \
          \langle \alpha|\lambda_1\lambda_2 \rangle \
      \langle \, \psb_{i\lambda_1} \, \psi_{i\lambda_2}
      \, \varphi_\beta \, \rangle_c \; ,
\label{opbpf}
\end{eqnarray}
\begin{eqnarray}
\partial_t \, \langle \, \pi_\alpha \, \pi_\beta \, \rangle_c
& = & \sum_{\lambda} \ t_{\alpha \lambda} \
      \langle \, \varphi_\lambda \, \pi_\beta \, \rangle_c
      \  + \:  t_{\beta \lambda} \
      \langle \, \pi_\alpha \, \varphi_\lambda \, \rangle_c
      \nonumber\\
& & - \: g_B \sum_{\lambda_1 \lambda_2} \
      \langle \alpha|\lambda_1\lambda_2 \rangle \
      \langle \, \psb_{i \lambda_1} \, \psi_{i \lambda_2}
      \, \pi_\beta \, \rangle_c
      \nonumber\\
& & - \: g_B \sum_{\lambda_1 \lambda_2} \
          \langle \beta|\lambda_1\lambda_2 \rangle \
      \langle \, \psb_{i \lambda_1} \, \psi_{i \lambda_2}
      \, \pi_\alpha \, \rangle_c \; ,
\label{opbpp}
\end{eqnarray}
\begin{eqnarray}
\partial_t \, \langle \, \psb_{a\alpha} \, \psi_{b\beta} \, \rangle_c
& = & \sum_{\lambda} \ \tilde{t}_{\lambda\alpha}^{la} \
      \langle \, \psb_{l\lambda} \, \psi_{b\beta} \, \rangle_c \
    - \: \tilde{t}_{\beta\lambda}^{bn} \
      \langle \, \psb_{a\alpha} \, \psi_{n \lambda} \, \rangle_c
      \nonumber\\
& &  + \:  i g_B \beta_{la} \sum_{\lambda_1 \lambda_2} \
      \langle \alpha|\lambda_1\lambda_2 \rangle \
      \langle \, \psb_{l \lambda_1} \, \psi_{b \beta}
              \, \varphi_{\lambda_2} \, \rangle_c
      \nonumber\\
& & - \: i g_B \beta_{bl} \sum_{\lambda_1 \lambda_2} \
      \langle \beta|\lambda_1\lambda_2 \rangle \
      \langle \, \psb_{a \alpha} \, \psi_{l \lambda_1}
             \, \varphi_{\lambda_2} \, \rangle_c \; ,
\label{opbwv}
\end{eqnarray}

Due to their length the equations of motion for further Green 
functions are shifted to Appendix B. 
A diagrammatical representation particularly for the
time evolution of a mixed 3-point function is given in 
Appendix C.

By using a finite basis of plane waves two cutoffs -- 
ultraviolet as well as infrared -- are introduced.
For a given boxsize $L$ the difference between two 
neighboring momenta is $\Delta k = 2\pi / L$ such that a 
basis of n states in $1+1$ dimensions contains an 
ultraviolet momentum-cutoff of $k_{cut} = (n-1) \Delta k / 2$.
\subsection{The renormalization scheme for Yukawa-theory}
\begin{figure}[h]
{\vspace*{5.0 cm} \hspace*{1.5 cm} 
{\rotate[l] {\psfig{file=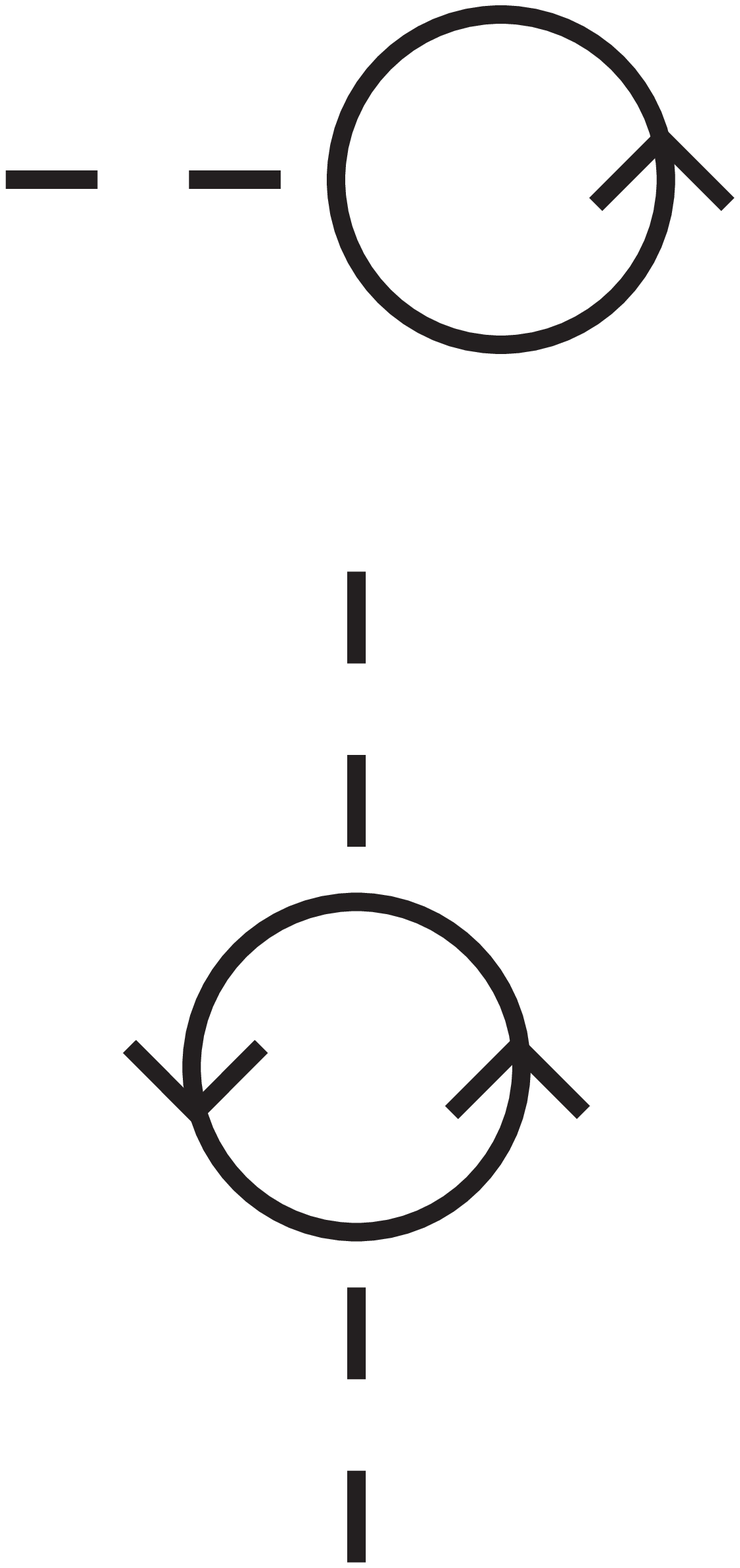,height=12 cm,width=8 cm}}}}
\caption{The divergent diagrams of 
Yukawa-theory in $1+1$ dimensions: 
the tadpole-diagram with superficial degree of divergence 
$D = 1$ (left) and the polarization-diagram with $D = 0$ (right).\label{bild1}}
\end{figure}

%
Before carrying out any calculations on the basis of the
correlation dynamical equations of motion one first has
to renormalize the Yukawa-theory.
In $1+1$ dimensions the theory contains two divergent
1PI-diagrams, the tadpole diagram and the polarization
diagram (cf. Fig.~\ref{bild1}), as can be seen directly 
by evaluating the superficial degree of divergence.

The tadpole diagram -- due to its structure with only one 
external boson line -- requires a renormalization of the 
external boson source.
Therefore we have to consider a Yukawa-Lagrangian extended 
by an additional boson source term $J_B \, \Phi$.
The renormalization is then done by introducing a source
counterterm which cancels exactly the divergences
arising from the tadpole diagram.
In contrast to this prescription the bare 
source $J_B$ is fixed in the GEP approach by the requirement 
that the energy density has to have a extremum at $\Phi_0 = 0$ 
(cf. \cite{shr86}).
The resulting difference between the unrenormalized and the 
renormalized source in GEP calculation is fully equivalent 
to the counterterm of the diagrammatical source renormalization.
One has to keep in mind that this renormalization 
procedure for the boson source does not affect the 
correlation dynamical equations of motion of the 
connected equal-time Green functions.
This results from the fact that the boson source term 
only appears in the equations of motion
for the 1-point functions.
Since the 1-point functions are taken to be fixed by means
of the separation (\ref{sep1}) no adjustments in the 
integration procedure are necessary.

The handling of the divergent polarization diagram,
however, is more sophisticated.
It turns out that the only possibility to get 
rid of the divergence consists in a multiplicative 
renormalization of the coupling constant.
In the continuum the renormalized coupling constant has
to be defined as
\be
g_R^2 \: = \: g_B^2 \ I_0(M) \; ,
\label{coprencont}
\ee
where $I_0$ is an -- in 1+1 dimensions logarithmically
divergent -- integral of the type
\bea
I_n(\hat{m}) \: = \: \int \frac{dk}{2\pi}
                  \; \hat{\omega}^{2n-1}_k(\hat{m}) \; ,
\qquad
\hat{\omega}_k(\hat{m}) =\sqrt{{\vec k}^2 + \hat{m}^2} \; .
\label{gepint}
\eea
Thus for a finite renormalized coupling constant the bare 
coupling has to be infinitesimally small.
We adopt the renormalization scheme of Ref. \cite{shr86}.

It is worth mentioning that there is no other renormalization
prescription possible.
A mass renormalization, which is the naive approach to 
a diagram of that structure, is not viable due to the 
equivalence of the renormalized masses and the effective 
masses at $\Phi_0 = 0$ in the GEP-formulation.
As a consequence the renormalized and unrenormalized 
masses are the same for the bosons as well as for 
the fermions, respectively.
On the other hand, a renormalization of the wavefunctions 
does also not lead to a solution since the effective 
potential will become unbounded from below.

The GEP-calculation in the continuum leads to the following
approximation for the effective potential of the Yukawa-theory
in $1+1$ dimensions:
\bea
V^{cont}_{GEP} (\Phi_0) 
& = &
\left[ \frac{1}{2} m^2 - g^2_R \right] {\Phi_0}^2 \; .
\label{geprenkont}
\eea
In order to compare the GEP solution with a correlation 
dynamical calculation one has to modify this expression 
due to the evaluation of the theory in discretized 
momentum space.
In this case we remain with the following expression 
for a discretized GEP:
\bea
V^{dis}_{GEP} (\Phi_0) \
& = & 2 \; S_1(M) \ - \ 2 \ S_1(M + g_B \Phi_0)
+ \ 2 M g_B \Phi_0 \ S_0(M)
      \ + \ \frac{1}{2} m^2 {\Phi_0}^2 \; .
\label{gepdis}
\eea
In comparison to the continuum GEP the integrals are replaced
by the corresponding sums:
\bea
S_n(\hat{m}) & = & \frac{1}{L} \ \sum_{i} \ {\hat{\omega}^{2n-1}_{k_i}(\hat{m})} \; ,
\qquad
\hat{\omega}_{k_i}(\hat{m}) =\sqrt{{\vec k}_i^2 + \hat{m}^2} \; .
\label{gepsum}
\eea
We note that in the GEP as well as in any correlation
dynamical calculation the vacuum expectation values with
respect to the perturbative vacuum are subtracted.
Finally, the multiplicative renormalization procedure 
for the Yukawa coupling constant has to be modified 
in the same way:
\be
g^2_R \: = \: g^2_B \ S_0(M) \; .
\ee
\newpage
\section{Numerical results}
\subsection{2- and 3-point approximations}
We first investigate the structure of the effective potential 
of the Yukawa theory in $1+1$ dimensions in the lowest orders
of correlation dynamics, i.e. in 2-point and 
3-point approximation.
In these calculations the effective potential is generated via 
an adiabatic switching of the coupling constant $g_R$ in time. 
The coupling constant is increased linearly 
-- starting from the unperturbed groundstate with $g_R = 0$ --
according to $g_R/m = \alpha t$, where t denotes the propagation 
time and $\alpha$ the ``adiabaticity'' parameter.
In the limit $\alpha \rightarrow 0$ the Gell-Mann and Low
theorem guarantees that the groundstate of the correlated
system for fixed coupling $g_R$ is reached \cite{gml51}.
For a detailed discussion of the applicability of this method 
we refer to Ref. \cite{phc97}, where this has been investigated
explicitly for $\Phi^4$-theory.
In the present case we only show that the applicability 
is given for all couplings in the correlation dynamical
2-point approximation.
For the calculations we use $m=M=5$ MeV while the system
of equations is integrated in a basis of 19 plane waves 
and a boxsize $L=100$ fm.
  
\begin{figure}[htbp]
{\vspace*{3.0 cm}}
{\psfig{file=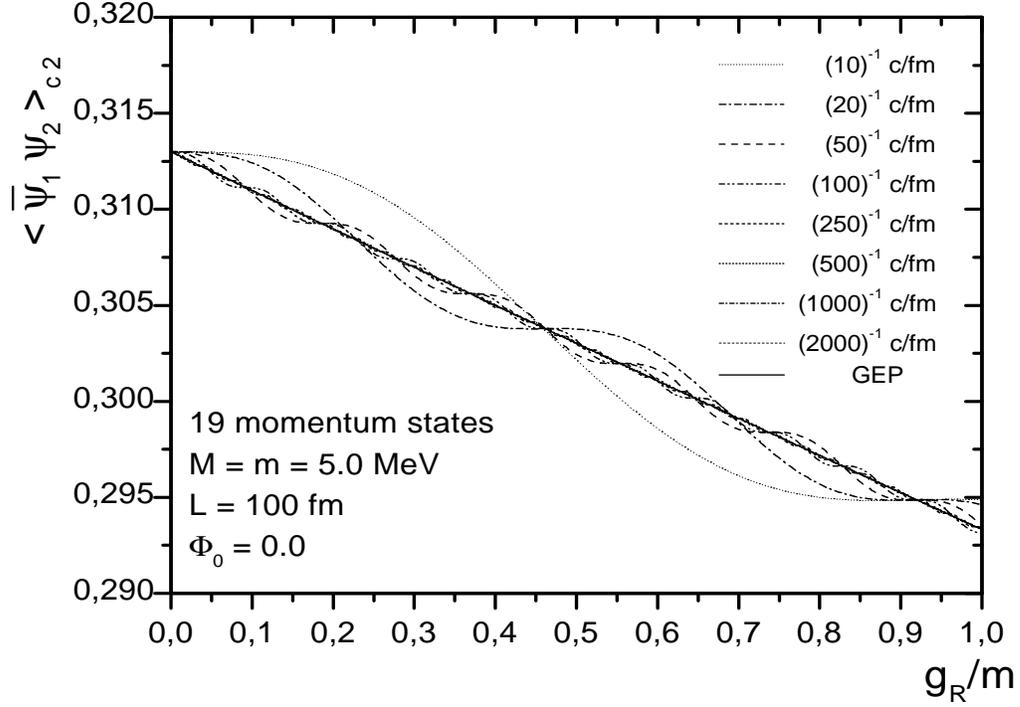,height=20.0cm,width=15.0cm}}
\vspace{-10.0 cm}
\caption{
One coefficient of the fermionic 2-point function 
$\langle \psb_1 \psi_2 \rangle_c$ as a function of the
dimensionless renormalized coupling constant $g_R/m$ for
various adiabaticity parameters $\alpha$.
For $\alpha \rightarrow 0$ one observes convergence of
the $Y_{1+1}CD(2)$-results towards the GEP-solution
(solid line).
\label{bild2}}
\end{figure}

To show the adiabatic convergence we display the real part 
of one coefficient of the connected Green function 
$\langle \psb_1 \psi_2 \rangle_c$
in $Y_{1+1}CD(2)$-approximation evaluated for various 
parameters $\alpha$ from $\alpha = 10^{-1}$ c/fm to
$\alpha = 2000^{-1}$ c/fm for a bosonic field expectation
value $\Phi_0 = 0$ (cf. Fig.~\ref{bild2}). 
With decreasing $\alpha$ one finds a reduction of the
oscillation amplitudes.
The corresponding increase of the oscillation frequency
is a consequence of the presentation as a function of the
dimensionless coupling $g_R/m$ and therefore an artefact
of the time rescaling.
For $\alpha \le 500^{-1}$ c/fm the calculations have 
sufficiently well converged versus an asymptotic curve.
All values of the observed quantity agree very well with a 
corresponding discretized GEP-calculation as presented by 
the solid line in Fig.~\ref{bild2}.
This also holds for all other coefficients of 
$\langle \psb_1 \psi_2 \rangle_c$.
Furtheron -- except explicitly mentioned otherwise -- we use
$\alpha = 2000^{-1}$ c/fm which ensures convergence in all
coupling regimes investigated.
In 3-point and 4-point approximation, however, this is no
longer possible above some critical coupling (see below).
   
In Fig.~\ref{bild3} the effective potential 
of the Yukawa-theory in $1+1$ dimensions is presented
in $Y_{1+1}CD(2)$- and for comparison in discretized 
GEP-approximation -- as a function of the dimensionless
coupling constant $g_R/m$ and the boson field 
expectation value $\Phi_0$.
In both approximations one recognizes the parabolic
structure of the effective potential for small
renormalized coupling constants.
With increasing coupling this parabola bends over and the 
effective potential turns negative above some critical coupling.
In contrast to the pure parabolic structure of the continuum
GEP (even for large $g_R/m$) we obtain a double-well shape
in the discretized calculation.
This behaviour is a direct consequence of the 
ultraviolet cutoff induced by the finite number of
single-particle basis states and can be explained by a
Taylor-expansion of the discretized effective potential
(\ref{gepdis}) around $\Phi_0 = 0$,
\bea
V^{dis}_{GEP} (\Phi_0) \
& = & \left[ \: \frac{1}{2} m^2 - g^2_R + g^2_B M^2 S_{-1}(M) \: \right]
      \Phi^2_0
\\
& & + \; g^3_B \Phi^3_0
        \left[ \: \frac{}{} M S_{-1}(M) - M^3 S_{-2}(M) \: \right]
      \nonumber\\
& & + \; g^4_B \Phi^4_0
        \left[ \: \frac{1}{4} S_{-1}(M)
             - \frac{3}{2} M^2 S_{-2}(M)
             + \frac{5}{4} M^4 S_{-3}(M)       \: \right]
\; + \; O(\Phi^5_0) \; . \nonumber
\label{geptaylor2}
\eea
Due to the ultraviolet cutoff the bare coupling constant
does not become infinitesimally small but assumes a finite
value.
Therefore, in addition to the terms in the continuum-limit,
further contributions of higher order in $\Phi_0$ appear.
Since the coefficients of the most important terms, the
$S_{-1}$-terms, are positive, this leads to an increase of
the GEP for high values of $| \Phi_0 |$.

Another feature of the correlation dynamical approximation
on the 2-point-level within a finite set of basis states
is the fact that for certain parameter regions the
$Y_{1+1}CD(2)$-approximation differs from the discretized
GEP-calculation.
However, in the continuum limit with an infinite number of
basis states both approximations generate the same results
since the GEP represents the stationary limit of the 
field-theoretical Hartree-Fock-Bogoliubov-(HFB)-approximation,
which is equivalent to the correlation dynamical
2-point approximation.
The difference in the discretized theory appears
in all areas of the parameter plane where the effective
fermion mass of the GEP-calculation
$\bar{M} = M + g_B \, \Phi_0$ becomes negative and can
be interpreted as a result of a level-crossing effect.
Beyond this hyperbola in the $g_R/m$-$\Phi_0$-plane
the fermion single-particle state of lowest energy in the 
noninteracting case, i.e. the zero momentum mode, is 
energetically preferred in comparison to the highest
lying fermion sea-state. 
Usually the highest sea-state and the first particle-state
in Dirac's-picture are separated by twice the effective
mass of the particle.
In our case, where the effective mass becomes less than
zero, this leads to an interchange of these states.
Since in the $Y_{1+1}CD(2)$-approximation no pair-production 
in the zero momentum mode is possible within the adiabatic 
process, it does not reproduce the true groundstate 
on the 2-point-level in the parameter-regions where
the level-crossing appears. 

\begin{figure}[t]
\vspace*{-2.0cm}
\hspace*{1.0cm}
{\psfig{file=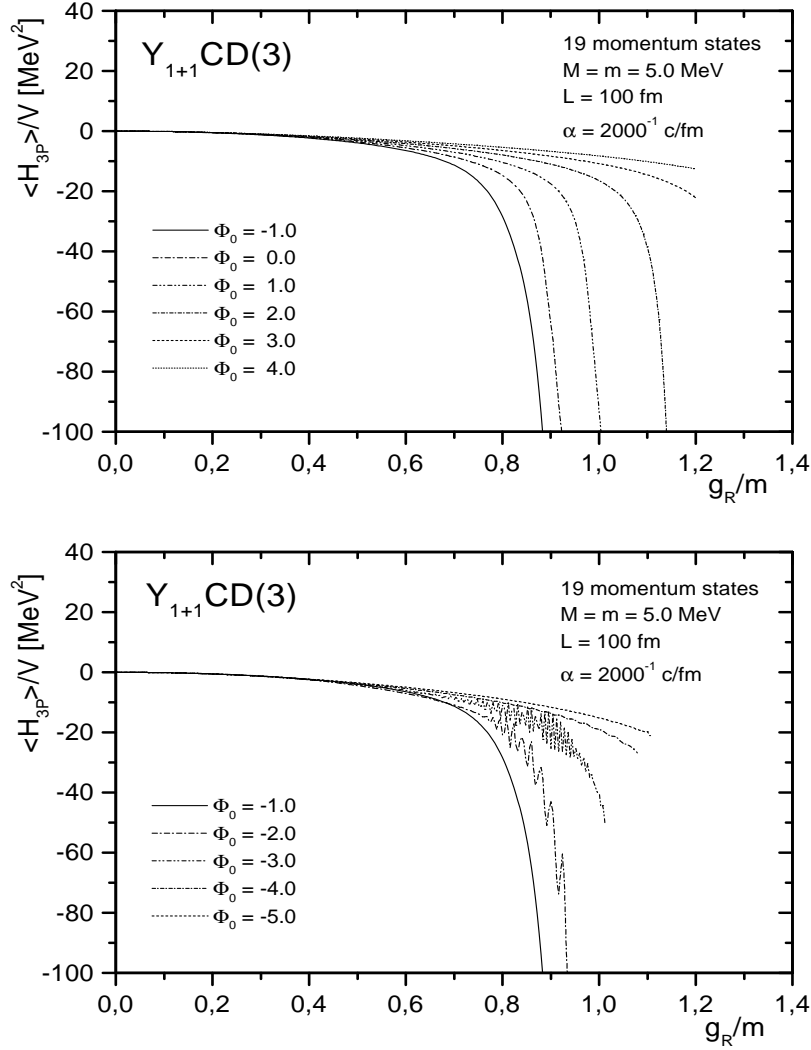,height=16.0 cm,width=12.0 cm}}
\caption{
3-point contribution to the total energy density for
various values of the boson background field $\Phi_0$
in $Y_{1+1}CD(3)$-calculation;
for $\Phi_0 \ge -1.0$ (upper part) and 
for $\Phi_0 \le -1.0$ (lower part).
The propagation breaks down for $\Phi_0$ in the non-convex 
region of the effective potential.
As ``breakdown point'' we specify those couplings at
which either the propagation collapses or the energy-density
of the 3-point coupling part starts to decrease rapidly.
\label{bild4}}
\end{figure}

In the $Y_{1+1}CD(3)$-approximation we find two new
features in comparison to calculations on the 2-point-level.
The first one is the existence of coupling regimes for small
boson background fields where the propagation of the system
of equations of motion breaks down.
Thus results for the effective potential in 
$Y_{1+1}CD(3)$-approximation are only accessible up to coupling 
constants of $g_R/m \approx 0.8$ for all $\Phi_0$. 
In contrast, the effective potential in 2-point approximation 
can be evaluated for all values of the coupling and the external 
background field.
One finds that the correlation dynamical propagation breaks
down as soon as the effective potential assumes a non-convex
structure (cf. Fig.~\ref{bild4}) for the set of parameters
($g_R/m$,$\Phi_0$).
This effect is known from the correlation dynamical 
investigations of $\Phi^4$-theory and was studied
in Ref. \cite{phc97}.
The reason for the breakdown lies in the fact that the cluster
decomposition property of connected Green functions is not valid
anymore in a two-phase configuration as encountered in the
``non-convex'' region of an effective potential, where the correct 
energy density has a upper bound given via a Maxwell-construction
\cite{zju89,ps95} that can be reached dynamically as soon
as the approximation scheme allows for tunneling.
Due to the inadequacy of the cluster decomposition the
connected Green functions are not restricted to small values,
but can dominate the full Green functions.
In the latter case the correlation dynamical truncation is no longer
a valid approximation.
It is worthwhile to point out that this unstable behaviour
in the present case is not a weakness of the correlation
dynamical approach since for those coupling strengths where
the discretized effective potential becomes non-convex, the system
is no longer bounded from below in the continuum.

\begin{figure}[t]
{\psfig{file=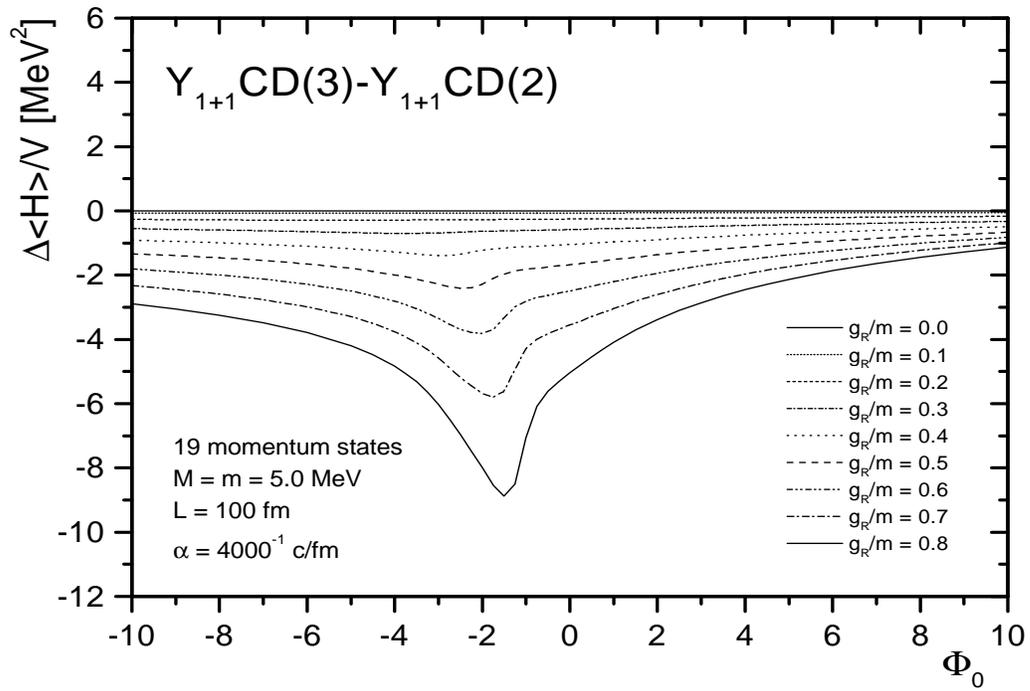,height=20.0 cm,width=15.0 cm}}
\vspace*{-10.0 cm}
\caption{
Difference of the effective potential in
$Y_{1+1}CD(3)$- and in $Y_{1+1}CD(2)$-approximation.
\label{bild5}}
\end{figure}

The second aspect of the $Y_{1+1}CD(3)$-approximation
is the reduction of the total energy density. 
While the global shape of the effective potential -- up to 
the critical coupling constant -- is not changed much one 
finds a significant lowering of the energy density especially 
for boson background fields with small absolute values 
(cf. Fig.~\ref{bild5})).
This effect shows its maximum influence not for $\Phi_0 = 0$,
but for such combinations of the coupling constant and 
the external background field where the $Y_{1+1}CD(2)$-approximation
separates from the discretized GEP, i.e. where the effective
fermion mass becomes zero.

The reason for this behaviour is that in 
the $Y_{1+1}CD(3)$-approximation (in contrast to the
$Y_{1+1}CD(2)$-approximation) pair-production in the
fermionic zero-momentum mode can take place. 
Thus the energy density can be lowered significantly
in a region where such a state is energetically preferred.
In order to understand why pair production in the fermionic
zero mode is not possible in the $Y_{1+1}CD(2)$-approximation,
we treat the problem in particle-number representation.
We consider the mode expansions (within a discretized 
momentum space) for all field operators
\begin{eqnarray}
\Phi({\vec x}) \; = \; 
\frac{1}{\sqrt{L}} \, \sum_{\alpha} \, \sqrt{2\omega_{\alpha}} \,
  \left[ \, a^{\phantom{\dagger}}_{\alpha} \, +  
         \, a^{\dagger}_{\bar{\alpha}}     \,   \right]
  {\rm e}^{i \vec{k}_{\alpha} \vec{x}} \; ,
\qquad
\qquad
\; \; \; \; \; \; \; \; \; 
\omega_{\alpha} =\sqrt{\vec{k}^2_{\alpha} + m^2} \: ,
\end{eqnarray}
\begin{eqnarray}
\Pi({\vec x}) \; = \; 
\frac{-i}{\sqrt{L}} \, \sum_{\alpha} \, \sqrt{\frac{\omega_{\alpha}}{2}} \,
  \left[ \, a^{\phantom{\dagger}}_{\alpha} \, -  
         \, a^{\dagger}_{\bar{\alpha}}     \,   \right]
  {\rm e}^{i \vec{k}_{\alpha} \vec{x}} \; ,
\end{eqnarray}
\begin{eqnarray}
\Psi_i({\vec x}) \; = \; 
\frac{1}{\sqrt{L}} \, \sum_{\alpha} \, \sqrt{\frac{M}{\Omega_{\alpha}}} \,
  \left[ \, b^{\phantom{\dagger}}_{\alpha} \, u_{i \alpha} \, +  
         \, d^{\dagger}_{\bar{\alpha}}     \, v_{i \bar{\alpha}} \, \right]
  {\rm e}^{i \vec{k}_{\alpha} \vec{x}} \; ,
\qquad \quad \; \;
\Omega_{\alpha} =\sqrt{\vec{k}^2_{\alpha} + M^2} \: ,
\end{eqnarray}
\begin{eqnarray}
\Psb_i({\vec x}) \; = \; 
\frac{1}{\sqrt{L}} \, \sum_{\alpha} \, \sqrt{\frac{M}{\Omega_{\alpha}}} \,
  \left[ \, b^{\dagger}_{\bar{\alpha}} \, \bar{u}_{i \bar{\alpha}} \, +  
         \, d^{\phantom{\dagger}}_{\alpha}  \, \bar{v}_{i \alpha} \, \right]
  {\rm e}^{i \vec{k}_{\alpha} \vec{x}} \; ,
\end{eqnarray}
where each annihilation operator annihilates the perturbative
vacuum. \\
Here the abbreviation $\cdot_{\bar{\alpha}}$ for the inverse 
momentum index is introduced according to
\bea
\vec{k}_{\alpha} + \vec{k}_{\bar{\alpha}} 
& = & \vec{0} \quad \forall \alpha \; .
\eea
Evaluating the complete equation of motion for the fermionic
particle-number operator we find (using
$\check{a}_{\alpha} \, = \, a_{\alpha} + a^{\dagger}_{\bar{\alpha}}$) :
\begin{eqnarray}
i \partial_t \, \langle \, b^{\dagger}_{\alpha} \, b^{ }_{\alpha} \, \rangle
& = & \phantom{-} \,
g_B M L \sum_{\lambda_1 \lambda_2} \,
\langle 1 | \alpha \lambda_1 \lambda_2 \rangle \,
\frac{1}{\sqrt{2 \, \Omega_{\alpha} \,
                    \Omega_{\lambda_1} \,
                    \omega_{\lambda_2}}}
\nonumber\\
& & \qquad \qquad \times
\left\{
\langle \, b^{\dagger}_{\alpha} 
           b^{ }_{\bar{\lambda}_1} 
           \check{a}^{ }_{\bar{\lambda}_2} \, \rangle \,
\bar{u}^{ }_{i \alpha} u^{ }_{i \bar{\lambda}_1} \, + \,
\langle \, b^{\dagger}_{\alpha} 
           d^{\dagger}_{\lambda_1} 
           \check{a}^{ }_{\bar{\lambda}_2} \, \rangle \,
\bar{u}^{ }_{i \alpha} v^{ }_{i \lambda_1} \,
\right\}
\nonumber\\
& & - \,
g_B M L \sum_{\lambda_1 \lambda_2} \,
\langle 1 | \alpha \lambda_1 \lambda_2 \rangle \,
\frac{1}{\sqrt{2 \, \Omega_{\alpha} \,
                    \Omega_{\lambda_1} \,
                    \omega_{\lambda_2}}}
\nonumber\\
& & \qquad \qquad \times
\left\{
\langle \, b^{\dagger}_{\bar{\lambda}_1} 
           b^{ }_{\alpha} 
           \check{a}^{ }_{\lambda_2} \, \rangle \,
\bar{u}^{ }_{i \bar{\lambda}_1} u^{ }_{i \alpha} \, + \,
\langle \, d^{ }_{\lambda_1} 
           b^{ }_{\alpha} 
           \check{a}^{ }_{\lambda_2} \, \rangle \,
\bar{v}^{ }_{i \lambda_1} u^{ }_{i \alpha} \,
\right\} \; .
\label{eombb}
\end{eqnarray}
In (\ref{eombb}) only operator products with 3 creation$/$annihilation
operators lead to a nonzero r.h.s. (even for the zero momentum mode),
such that for all momenta a particle-antiparticle
production is possible.
In a 2-point calculation where the expectation values of 
the operator products are factorized, only the following 
terms survive, 
\begin{eqnarray}
\langle \, \bullet_{\alpha} \, \bullet_{\beta} \, \check{a}_{\gamma} \, \rangle
\: = \:
\langle \, \bullet_{\alpha} \, \bullet_{\beta} \, \rangle \,
\langle \, \check{a}_{\gamma} \, \rangle \;
\: = \:
\langle \, \bullet_{\alpha} \, \bullet_{\beta} \, \rangle \:
\sqrt{2 \, m \, L } \, \Phi_0 \, \delta_{\vec{k}_{\gamma},\vec{0}} \; .
\label{factorized}
\end{eqnarray}
In (\ref{factorized}) the symbols $\bullet_{\alpha}$ represent 
an arbitrary fermionic operator with the corresponding momentum 
index.
In a translationally invariant system one ends up with  
\bea
\partial_t \, \langle \, b^{\dagger}_{\alpha} \, b^{ }_{\alpha} \, \rangle
 & = &
 - \, 2 \, g_B \, \Phi_0 \, \frac{k_{\alpha}}{\Omega_{\alpha}} \,
 Im
 \left\{
 {\langle \, b^{\dagger}_{\alpha} d^{\dagger}_{\bar{\alpha}} \, \rangle}
 \right\} \; 
 \label{eombbend}
\eea
which becomes zero for the zero momentum mode.
Thus when propagating the system in time -- even with a 
simultaneous variation of the coupling constant -- the 
explicit appearance of the momentum $k_{\alpha}$
in (\ref{eombbend}) prevents a change of the occupation number
in the zero momentum mode.
\subsection{The 4-point approximation}
The highest order correlation dynamical approximation investigated
in the present work is the $Y_{1+1}CD(4)$-approximation.
For these calculations we modify the size of the box to 5 fm
due to the enormous increase in the number of differential
equations that have to be integrated. 
In this approximation all connected boson 
3-point functions as well as all connected 
4-point functions, that contain fermion degrees of freedom,
must be propagated.
(Due to the special structure of the Yukawa interaction
the pure bosonic connected Green functions of order 4
disappear in this approximation scheme.)
For practical reasons we are limited to 11 basis states.
One has to choose a smaller boxsize in order to achieve a 
sufficiently good ultraviolet convergence 
(at least for relative small boson background fields and 
up to coupling constants of $g_R / m = 0.8$).
All other parameters, that have been used in the previous
calculations, remain the same.

\begin{figure}[t]
{\psfig{file=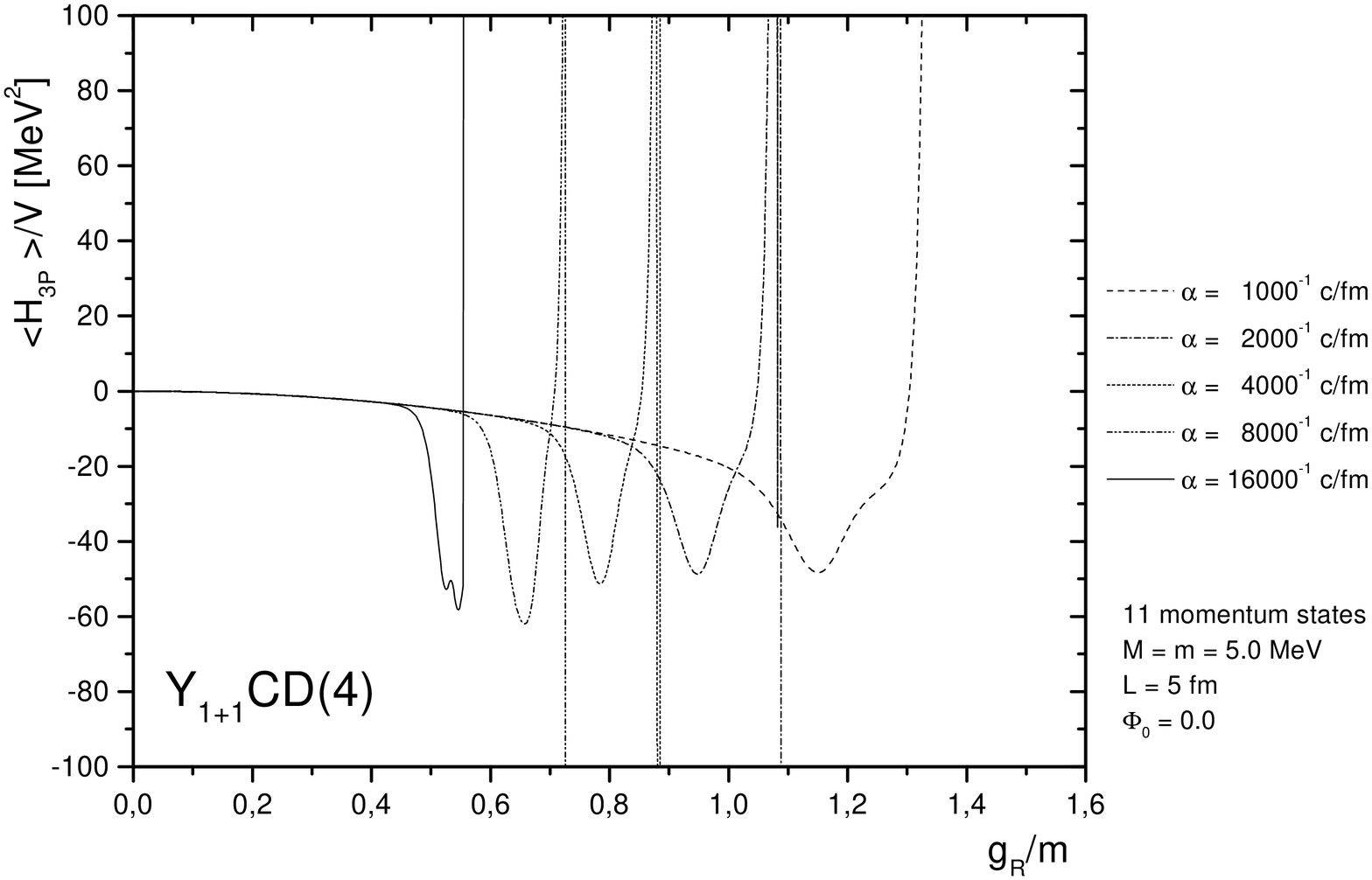,height=12.0 cm,width=16.0 cm}}
\caption{
3-point contribution to the energy density
in $Y_{1+1}CD(4)$-approximation for $\Phi_0 = 0$
as a function of the dimensionless coupling constant
$g_R/m$ for various adiabaticity parameters $\alpha$.
\label{bild6}}
\end{figure}

\begin{figure}[t]
{\psfig{file=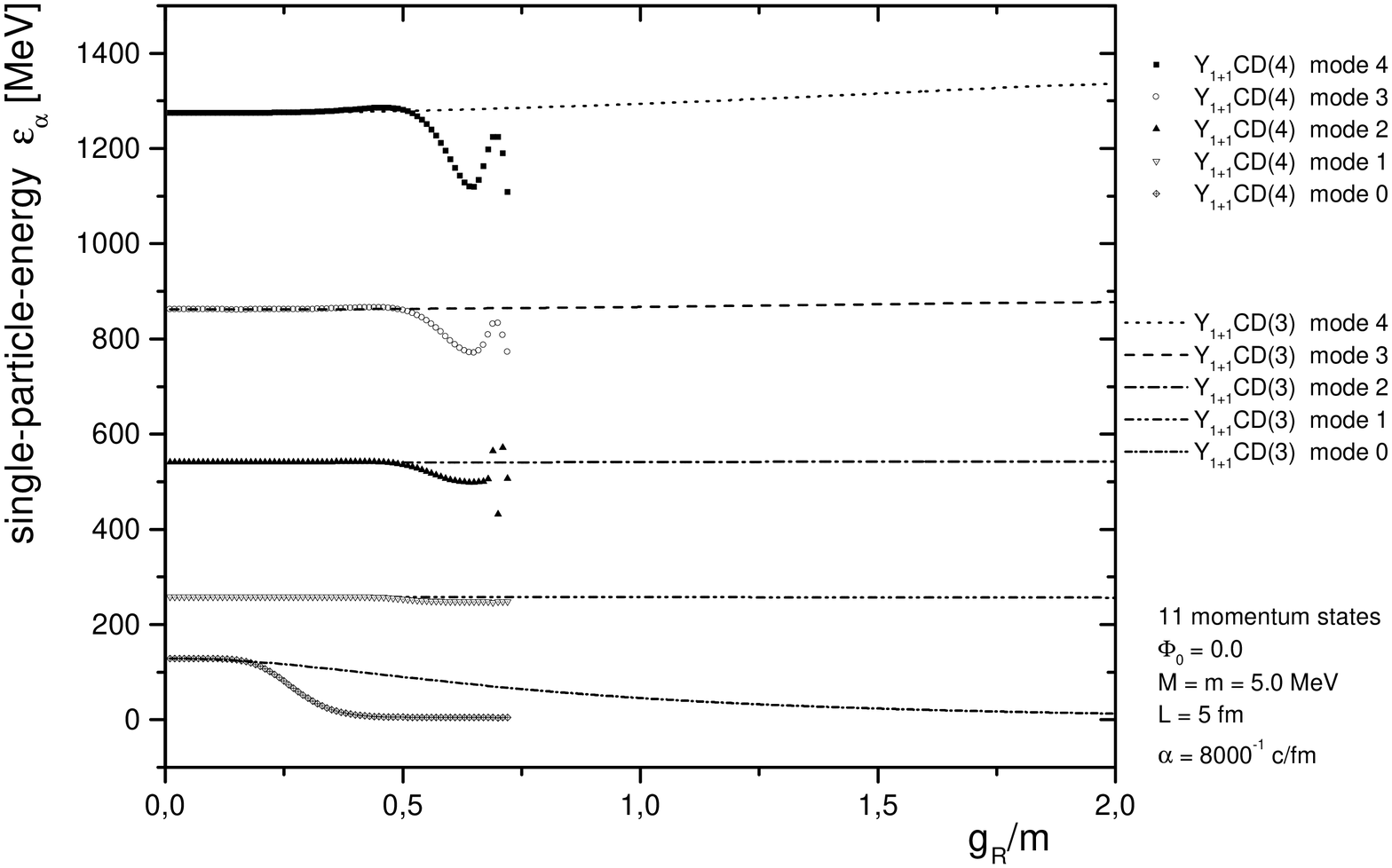,height=12.0 cm,width=16.0 cm}}
\caption{
Single-particle energies $\epsilon_{\alpha}$ for the
individual fermionic momentum modes in
$Y_{1+1}CD(3)$- and $Y_{1+1}CD(4)$-approximation
as a function of the dimensionless coupling constant
$g_R/m$.
Note the massive decrease in the zeroth mode
in the $Y_{1+1}CD(4)$-calculation at $g_R/m \approx 0.25$.
\label{bild7}}
\end{figure}

\begin{figure}[t]
{\psfig{file=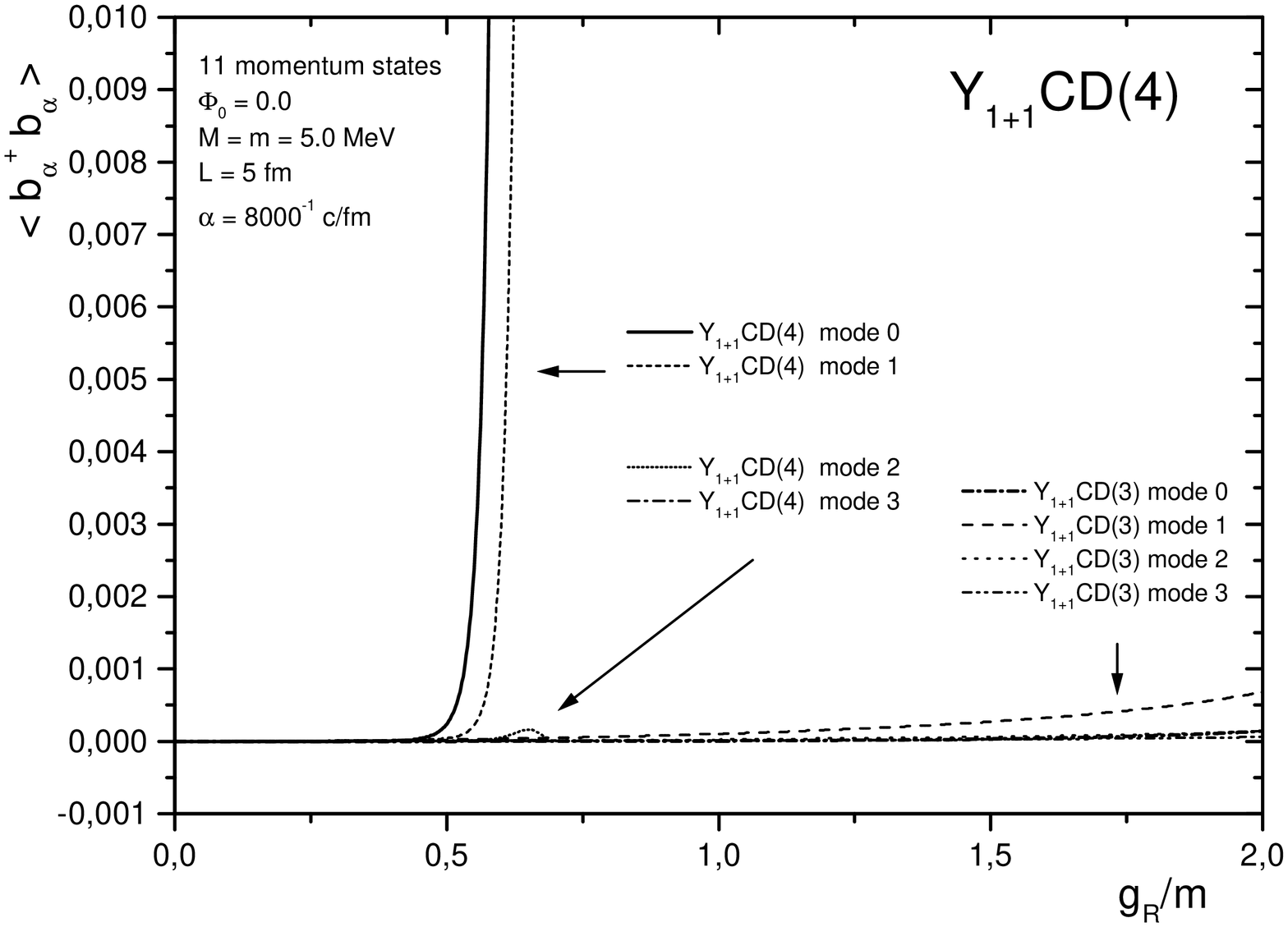,height=12.0 cm,width=16.0 cm}}
\caption{
The occupation numbers
$\langle b^{\dagger}_{\alpha} b^{\protect\phantom{\dagger}}_{\alpha} \rangle$
of the zeroth and first fermionic particle mode in
$Y_{1+1}CD(4)$-approximation as a function of $g_R/m$.
In the corresponding $Y_{1+1}CD(3)$-calculation the increase
is only moderate and the occupation number of the 
zero-momentum mode remains small even for strong couplings.
\label{bild8}}
\end{figure}

\begin{figure}[t]
{\vspace*{-1.5 cm}}
{\psfig{file=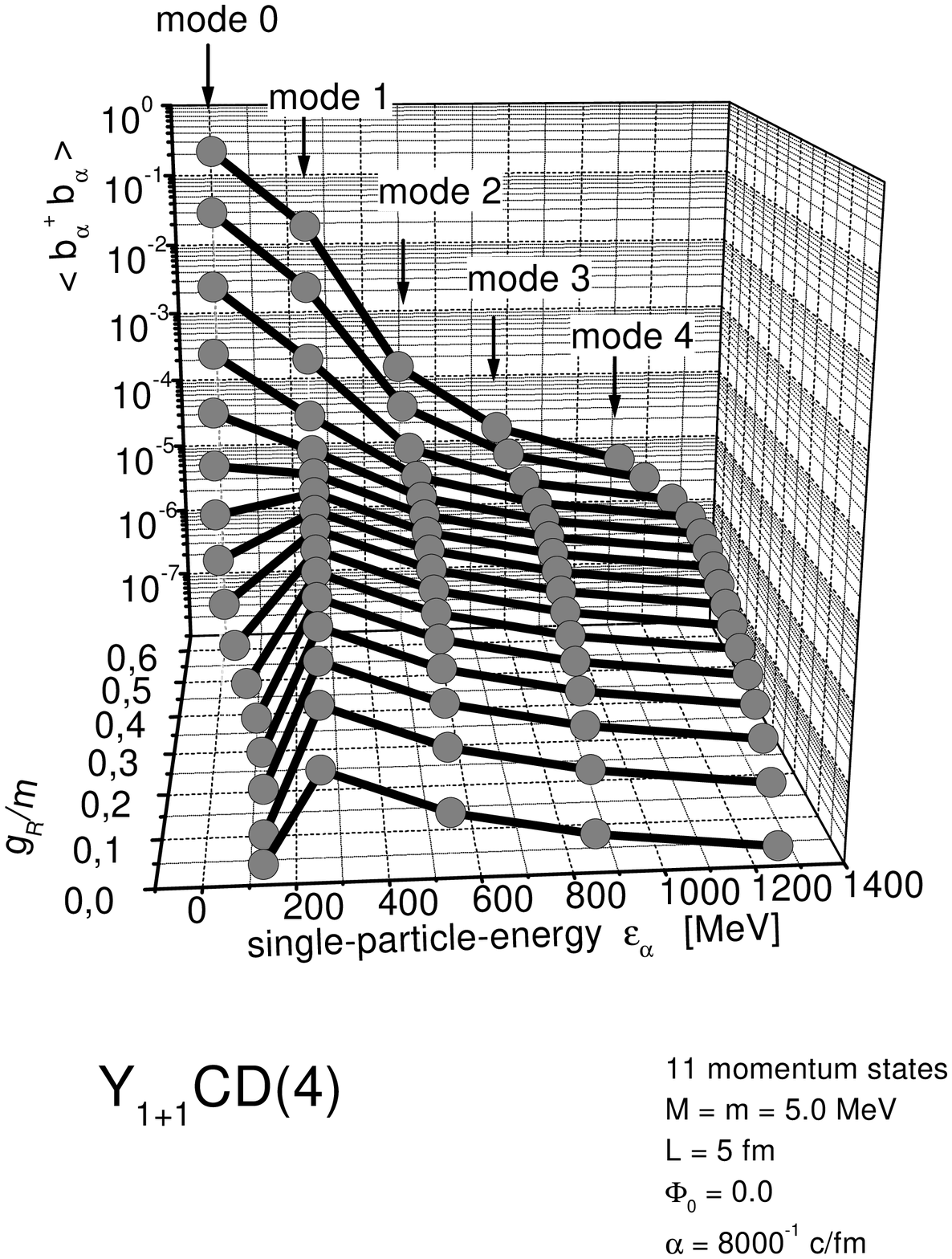,height=19.0 cm,width=15.0 cm}}
{\vspace*{-0.5 cm}}
\caption{
Fermionic particle occupation numbers
$\langle b^{\dagger}_{\alpha} b^{\protect\phantom{\dagger}}_{\alpha} \rangle$
of the individual momentum modes as a function of the
dimensionless coupling constant $g_R/m = 0.04 ... 0.64 \; (0.04)$
and the single-particle energy $\epsilon_{\alpha}$.
For coupling constants of about $\approx 0.25$
a strong reduction of the single-particle energies
$\epsilon_{\alpha}$ -- especially of the 
zero-momentum mode -- is observed.
\label{bild9}}
\end{figure}

In $Y_{1+1}CD(4)$-approximation we observe a novel effect  
for small boson background fields, i.e. the propagation
becomes unstable at $\Phi_0 = 0$ already for ``low'' coupling
constants $g_R/m < 0.4$.
In the limit $\alpha \rightarrow 0$ one no longer achieves 
adiabatic convergence for $g_R / m \ge 0.4$ (cf. Fig.~\ref{bild6}).
These coupling strengths are much lower than those for 
the non-convex phase of the system.
To analyse this behaviour in more detail we introduce the 
concept of an equal-time self energy.
One way to proceed this is to define a G-matrix 
$G^{ab}_{\alpha \beta \gamma}$ for the connected equal-time 
Green functions via the following relation between the
full and the unconnected parts of the mixed 3-point function:
\begin{eqnarray}
\sum_{\lambda_1 \lambda_2} \,
G^{la}_{\alpha \lambda_1 \lambda_2} \,
\langle \, \psb_{l \lambda_1} \, \psi_{b \beta} \, \rangle \,
\langle \, \varphi_{\lambda_2} \, \rangle
& = &
g_B \, {L}^{3/2} \, \gamma^0_{la} \,
\sum_{\lambda_1 \lambda_2} \,
\langle \, \alpha | \lambda_1 \lambda_2 \, \rangle \,
\langle \, \psb_{l \lambda_1} \, \psi_{b \beta} \,
\varphi_{\lambda_2} \, \rangle \: .
\label{gmdiff}
\end{eqnarray}
Due to translational invariance only the terms with 
$\vec{k}_{\lambda_2} = \vec{0}$ contribute on the l.h.s. 
of (\ref{gmdiff}) and the equation for the G-matrix reads
\begin{eqnarray}
\sum_{\lambda } \,
G^{la}_{\alpha \lambda} \,
\langle \, \psb_{l \lambda} \, \psi_{b \beta} \, \rangle \,
& = &
\frac{g_B \, L \, \gamma^0_{la}}{\Phi_0} \,
\sum_{\lambda_1 \lambda_2} \,
\langle \, \alpha | \lambda_1 \lambda_2 \, \rangle \,
\langle \, \psb_{l \lambda_1} \, \psi_{b \beta}
\, \varphi_{\lambda_2} \, \rangle \; ,
\label{gmeasy}
\end{eqnarray}
where 
\begin{eqnarray}
G^{ab}_{\alpha \beta}
& := &
\sum_{\gamma} \: G^{ab}_{\alpha \beta \gamma} \:
\delta_{\vec{k}_{\gamma},\vec{0}}
\end{eqnarray}
has been introduced.
Defining the self energy by
\begin{eqnarray}
\tilde{\Sigma}^{ab}_{\alpha \beta}
\; = \;
\frac{1}{\sqrt{L}} \,
\sum_{\lambda} \,
G^{ab}_{\alpha \beta \lambda} \,
\langle \, \varphi_{\lambda} \, \rangle
\; = \;
\Phi_0 \, G^{ab}_{\alpha \beta}
\label{selfdef} \; ,
\end{eqnarray}
we can write the equations of motion for the fermionic 
2-point function as
\begin{eqnarray}
i \partial_t \,
\langle \, \psb_{a \alpha} \, \psi_{b \beta} \, \rangle
& = &
\; i \,
\sum_{\lambda}
\left\{ \,
\tilde{t}^{la}_{\lambda \alpha} \,
\langle \, \psb_{l \lambda} \, \psi_{b \beta} \, \rangle
\, - \,
\tilde{t}^{bl}_{\beta \lambda} \,
\langle \, \psb_{a \alpha} \, \psi_{l \lambda} \, \rangle \,
\right\}
\nonumber\\
& & + \,
\sum_{\lambda}
\left\{ \,
\tilde{\Sigma}^{la}_{\lambda \alpha} \,
\langle \, \psb_{l \lambda} \, \psi_{b \beta} \, \rangle
\, - \,
\tilde{\Sigma}^{bl}_{\beta \lambda} \,
\langle \, \psb_{a \alpha} \, \psi_{l \lambda} \, \rangle \,
\right\} \; .
\end{eqnarray}
The self energies contain all contributions to the time
evolution that exceed the 2-point-level.
Keeping this in mind one can also define the
self energies directly (and omit the singularity at 
$\Phi_0 = 0$ in (\ref{gmeasy})), such that the interaction
part in the equation of motion has the same structure 
as the free contribution.
With regard to the observed behaviour
in $Y_{1+1}CD(4)$-approximation the self energy of the
expectation value of an operator product of
two fermionic particle operators is of interest:
\begin{eqnarray}
i \partial_t \, \langle \, b^{\dagger}_{\alpha} \, b^{ }_{\beta} \, \rangle
& = &
- \, \sum_{\lambda} \, \tilde{t}_{\lambda \alpha} \,
     \langle \, b^{\dagger}_{\lambda} b^{ }_{\beta} \, \rangle
\, + \,
     \sum_{\lambda} \, \tilde{t}_{\beta \lambda} \,
     \langle \, b^{\dagger}_{\alpha} b^{ }_{\lambda} \, \rangle
\nonumber\\
& &
+ \,
g_B M L \sum_{\lambda_1 \lambda_2} \,
\langle 1 | \beta \lambda_1 \lambda_2 \rangle \,
\frac{1}{\sqrt{2 \, \Omega_{\beta} \,
                    \Omega_{\lambda_1} \,
                    \omega_{\lambda_2}}}
\nonumber\\
& & \qquad \qquad \times
\left\{
\langle \, b^{\dagger}_{\alpha} 
           b^{ }_{\bar{\lambda}_1} 
           \hat{a}^{ }_{\bar{\lambda}_2} \, \rangle \,
\bar{u}^{ }_{i \beta} u^{ }_{i \bar{\lambda}_1} \, + \,
\langle \, b^{\dagger}_{\alpha} 
           d^{\dagger}_{\lambda_1} 
           \hat{a}^{ }_{\bar{\lambda}_2} \, \rangle \,
\bar{u}^{ }_{i \beta} v^{ }_{i \lambda_1} \,
\right\}
\nonumber\\
& & - \,
g_B M L \sum_{\lambda_1 \lambda_2} \,
\langle 1 | \alpha \lambda_1 \lambda_2 \rangle \,
\frac{1}{\sqrt{2 \, \Omega_{\alpha} \,
                    \Omega_{\lambda_1} \,
                    \omega_{\lambda_2}}}
\nonumber\\
& & \qquad \qquad \times
\left\{
\langle \, b^{\dagger}_{\bar{\lambda}_1} 
           b^{ }_{\beta} 
           \hat{a}^{ }_{\lambda_2} \, \rangle \,
\bar{u}^{ }_{i \bar{\lambda}_1} u^{ }_{i \alpha} \, + \,
\langle \, d^{ }_{\lambda_1} 
           b^{ }_{\beta} 
           \hat{a}^{ }_{\lambda_2} \, \rangle \,
\bar{v}^{ }_{i \lambda_1} u^{ }_{i \alpha} \,
\right\} \; . \\
\nonumber
\label{eombb2}
\end{eqnarray}
Defining the self energy directly by
\begin{eqnarray}
\tilde{\Sigma}_{\beta \lambda}
& = &
\frac{g_B \, M \, L }
{\langle \, b^{\dagger}_{\alpha} \, b^{ }_{\lambda} \, \rangle} \,
\sum_{\lambda_1 \lambda_2} \,
\langle 1 | \beta \lambda_1 \lambda_2 \rangle \,
\frac{1}{\sqrt{2 \, \Omega_{\beta} \,
                    \Omega_{\lambda_1} \,
                    \omega_{\lambda_2}}}
\nonumber\\
& & \qquad \qquad \times
\left\{
\langle \, b^{\dagger}_{\alpha} 
           b^{ }_{\bar{\lambda}_1} 
           \hat{a}^{ }_{\bar{\lambda}_2} \, \rangle \,
\bar{u}^{ }_{i \beta} u^{ }_{i \bar{\lambda}_1} \, + \,
\langle \, b^{\dagger}_{\alpha} 
           d^{\dagger}_{\lambda_1} 
           \hat{a}^{ }_{\bar{\lambda}_2} \, \rangle \,
\bar{u}^{ }_{i \beta} v^{ }_{i \lambda_1} \,
\right\} \; ,
\label{selfeinf}
\end{eqnarray}
we remain with the following expression for the 
equation of motion (\ref{eombb2}):
\begin{eqnarray}
i \partial_t \, \langle \, b^{\dagger}_{\alpha} b^{ }_{\beta} \, \rangle
& = &
- \, \sum_{\lambda} \,
     \left[ \,
     \tilde{t}_{\lambda \alpha} \,
   + \tilde{\Sigma}_{\lambda \alpha} \,
     \right] \,
     \langle \, b^{\dagger}_{\lambda} b^{ }_{\beta} \, \rangle
\, + \,
     \sum_{\lambda} \,
     \left[ \,
     \tilde{t}_{\beta \lambda} \,
   + \tilde{\Sigma}_{\beta \lambda} \,
     \right] \,
     \langle \, b^{\dagger}_{\alpha} b^{ }_{\lambda} \, \rangle \; ,
\end{eqnarray}
in which the free propagation part is given by
the energy in the noninteracting case, i.e.
$\tilde{t}_{\alpha \beta} = \Omega_{\alpha} \delta_{\alpha \beta}$.
With the help of these self energies one can construct
the renormalized mean-field
\begin{eqnarray}
\tilde{U}^{ren}_{\alpha \beta}
& = &
(Re \tilde{\Sigma})_{\alpha \beta} \; 
\end{eqnarray}
and the renormalized mean-field-hamiltonian
\begin{eqnarray}
\tilde{h}^{ren}_{\alpha \beta}
& = &
\tilde{t}_{\alpha \beta} \, + \,
\tilde{U}^{ren}_{\alpha \beta} \; .
\label{renmfh}
\end{eqnarray}
The eigenvalues $\epsilon_{\alpha}$ of (\ref{renmfh}) 
are the interesting quantities because they correspond 
to the energies of the basis states in the interacting 
case.
Plotting the energies $\epsilon_{\alpha}$ versus the
dimensionless coupling constant $g_R/m$ we obtain the 
following picture (Fig.~\ref{bild7}):
Whereas the single-particle energies do not change much
for $g_R / m \le 0.2$, at $g_R / m \approx 0.25$
a substantial lowering of the energy-levels -- especially
in the zero-momentum mode -- appears.
For all other momentum modes this behaviour is not so strong
and shifted to higher couplings $g_R/m$.
This contrasts the situation in the $Y_{1+1}CD(3)$-approximation.
In the correlation dynamical 3-point approximation
the decrease of the single-particle energies is much 
slower.
Here we have to mention that for this behaviour in the
$Y_{1+1}CD(4)$-calculation no -- in comparison to
the $Y_{1+1}CD(3)$-approximation -- additional
terms appear in the self energy.
Only the magnitude of the terms in the self energy
(\ref{selfeinf}) is changed significantly by the inclusion
of the 4-point functions.
Considering the occupation numbers 
$\langle b^{\dagger}_{\alpha} b^{ }_{\alpha} \rangle$
one finds -- as a consequence of this rapid decrease of the
single-particle energies -- a very strong population of the
corresponding modes for $g_R/m \approx 0.5$ in the
$Y_{1+1}CD(4)$-approximation (cf. Fig.~\ref{bild8}).
For small couplings the zero momentum mode
is occupied only weakly while the occupation number has
its maximum in the first mode; all other modes follow
corresponding to their single-particle energies.
When reaching the critical coupling regime the
single-particle energy for the zero-momentum mode
drops substantially such that it can be populated
to a high degree.
Finally, for increasing $g_R/m$ the occupation
number of this mode exceeds those of all other
momentum modes and the zero mode becomes occupied
preferentially (cf. Fig.~\ref{bild9}).
Here we also find a different picture in the
$Y_{1+1}CD(3)$-approximation, where the lowest
momentum modes are populated only for very high
coupling constants and with much less intensity
(cf. Fig.~\ref{bild8}).

\begin{figure}[t]
\hspace*{1.0cm} 
{\psfig{file=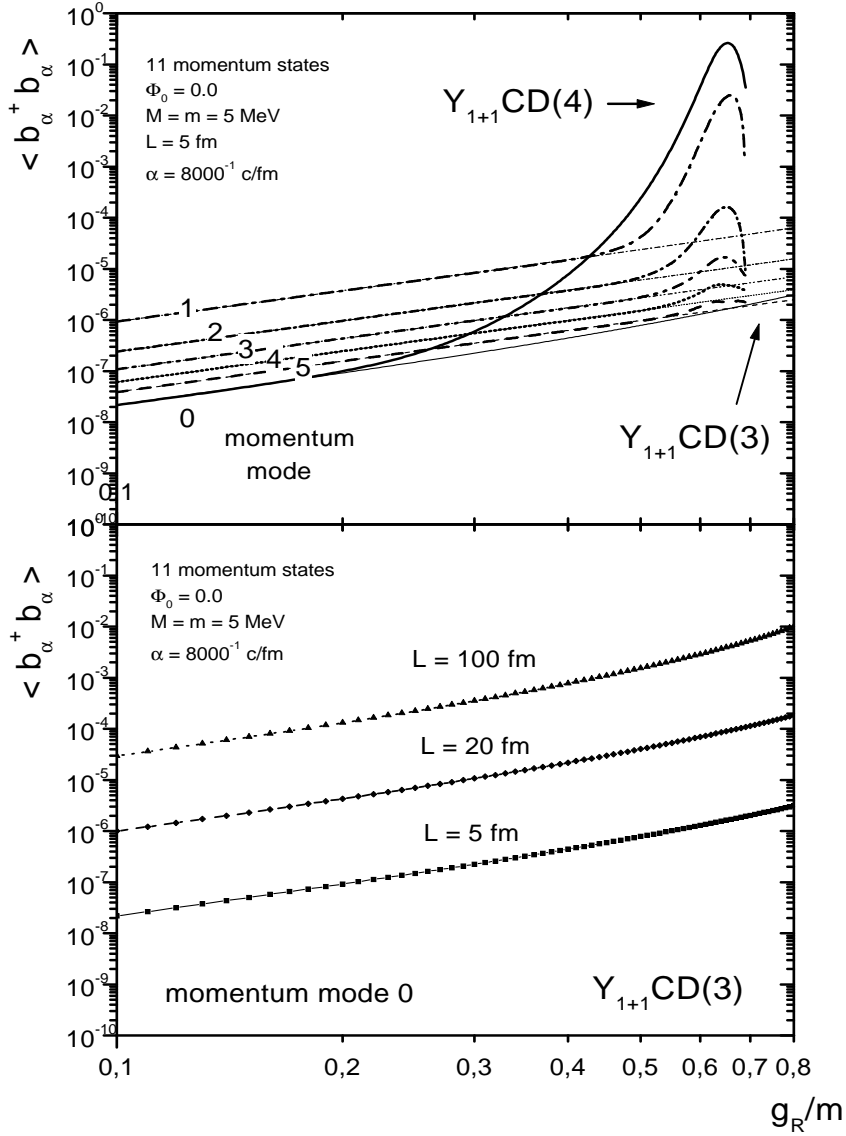,height=16.0 cm,width=12.0 cm}}
\caption{
Occupation numbers of the individual fermionic particle modes 
in $Y_{1+1}CD(3)$- and in $Y_{1+1}CD(4)$-approximation in the 
small coupling regime in double logarithmic representation
(upper part).
Occupation number of the lowest fermionic particle 
mode in $Y_{1+1}CD(3)$-approximation for various boxsizes
(L = 5 fm, 20 fm, 100 fm) (lower part).
\label{bild10}}
\end{figure}

The same behaviour can be seen in the alternative
representation in Fig.~\ref{bild10} (upper part), where 
the occupation numbers of the modes as well as the
dimensionless coupling constant are displayed
logarithmically.
For small coupling constants the occupation numbers
of all momentum modes are almost the same in both
approximations.
The linear increase indicates a scaling of the
occupation numbers with approximately the square of the
(unrenormalized) coupling constant.
We recall that, with exception of the zero-momentum
mode, the occupation numbers of all other modes in
$Y_{1+1}CD(3)$-approximation agree quite well with
those of the 2-point calculations.
One can recover this dependence on $g_R/m$ analytically from 
the GEP-solution in the small coupling limit.
Furthermore, one finds a scaling of the occupation numbers
with $\Phi^2_0$ if the absolute value of $\Phi_0$ is small.
Thus for low $g_R/m$ the occupation numbers of all
momentum modes (except the zero mode) are mainly determined
by the 2-point contributions.
The particle number in the zeroth mode is only
influenced -- due to the missing property of pair-production
in the 2-point approximation -- by contributions of
higher order connected Green functions.
Consequently the zero mode (although preferred by
energy) is less populated than all residual modes.
Its relative occupation number is increased in the
$Y_{1+1}CD(3)$-approximation for growing coupling $g_R/m$.
In $Y_{1+1}CD(4)$-approximation we find approximately
the same particle number (compared to the
$Y_{1+1}CD(3)$-approximation) for small $g_R/m$.
For coupling constants $g_R / m \approx 0.2$
the influence of the 4-point contribution begins to
grow and first leads to an increase of the occupation
number of the zero mode, then of all other modes
with growing $g_R/m$.

We have to point out that the observed effect persists 
in the ultraviolet limit and is not an effect of the 
regularization.
This we conclude from the fact that the decrease of
the single-particle energies and the subsequent
increase of the occupation numbers appears in all
momentum modes.
We speculate that this new effect is due to condensation
of fermion-antifermion pairs that only appears on the 
4-point-level.

In order to demonstrate that the steep rise in the 
occupation number of the zero momentum mode is not an 
artefact of the lower boxsize $L$ we show in 
Fig.~\ref{bild10} (lower part) the occupation number of 
the lowest momentum mode as a function of $g_R/m$ for 
three boxsizes ($L$ = 5 fm, 20 fm, 100 fm) in 3-point 
approximation using the usual parameterset of 
the 4-point approximation.
One recognizes that for all boxsizes $L$ the occupation 
number of the zero momentum mode increases only smoothly 
with $g_R/m$ and no breakdown appears in the displayed 
coupling regime.
Without explicit representation we note that the 
occupation numbers for all higher lying modes in the
3-point approximation essentially show a straight line 
as in the upper part of Fig.~\ref{bild10}, however,
shifted as a function of $L$.
Therefore the difference between the $Y_{1+1}CD(4)$- and the 
$Y_{1+1}CD(3)$-approximation is substantial and not just an
effect of the different $L$ chosen in both approximations before. 

\begin{figure}[t]
{\psfig{file=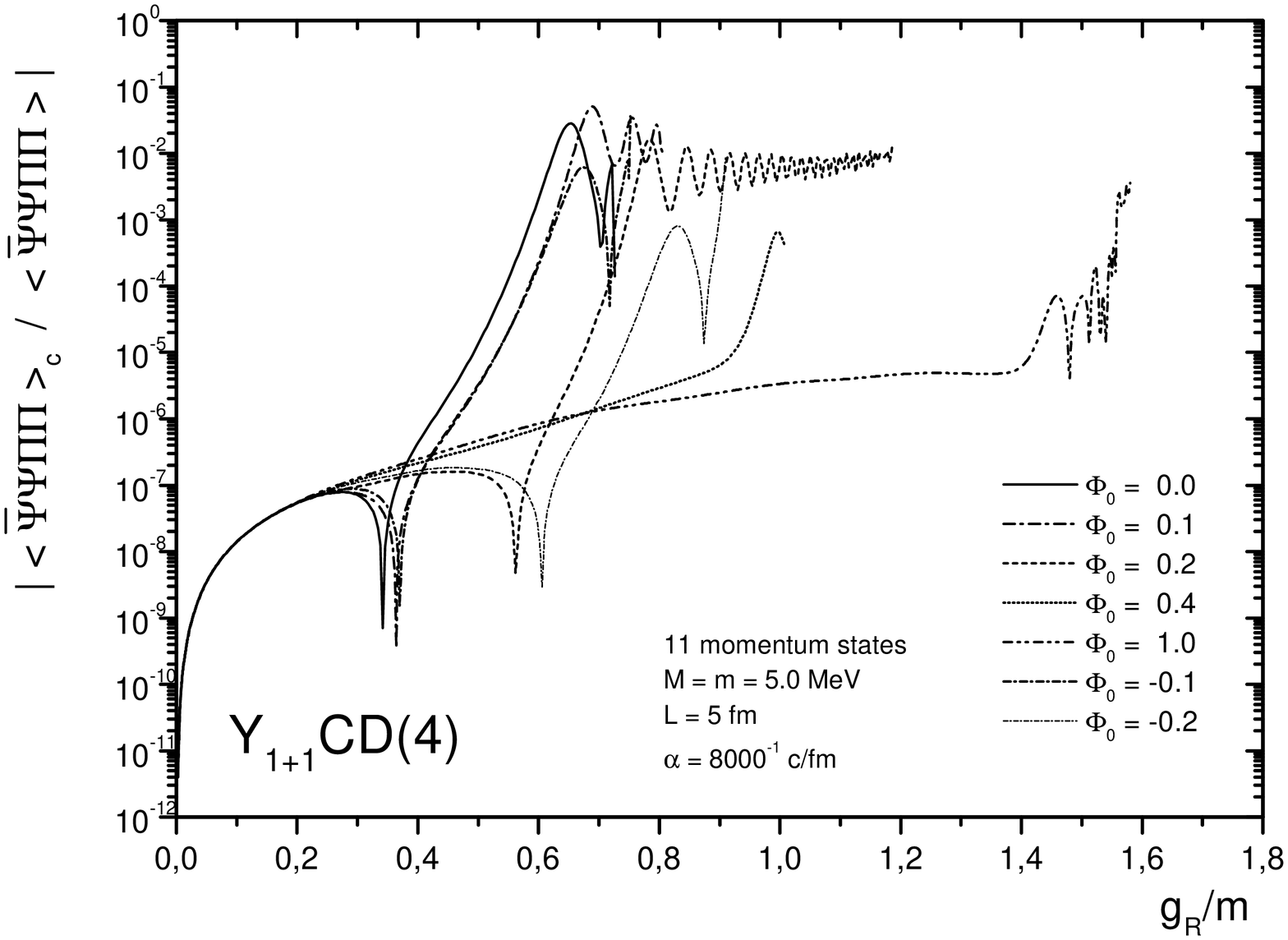,height=12.0 cm,width=16.0 cm}}
\caption{
Correlation strength
$| \langle \Psb \Psi \Pi \Pi \rangle_c \; / \;
   \langle \Psb \Psi \Pi \Pi \rangle | $
as a function of the dimensionless coupling constant
$g_R/m$ and the classical boson field $\Phi_0$.
\label{bild11}}
\end{figure}

The regime of the classical boson field $\Phi_0$,
where this behaviour can be found, is very small.
Even for $| \Phi_0 | \approx 1.0$ higher quantum
correlations are suppressed so that the effect is
no longer present.
To illustrate the dependence on $\Phi_0$ we show
in Fig.~\ref{bild11} the correlation strength of
the mixed 4-point-function 
$\langle \Psb \Psi \Pi \Pi \rangle$,
i.e. the quantity
\bea
\left| \: \frac{ \langle \Psb \Psi \Pi \Pi \rangle_c }
       { \langle \Psb \Psi \Pi \Pi \rangle   } \: \right| \; ,
\label{cswvpp}
\eea
(which plays an important role for the applicability
of the correlation dynamical approach \cite{phc97}) as 
a function of $g_R/m$ for various $\Phi_0$.
While the correlation strength increases smoothly
for small couplings, it grows rapidly for a
-- $\Phi_0$ dependent -- critical coupling.
Sharply peaked minima are a result of a change in
sign in (\ref{cswvpp}).
As discussed in detail in \cite{phc97} a high correlation
strength is allways connected with an increasing
instability of the propagation which in the present case
leads to a breakdown.
Here we find that the stability of the system depends
strongly on the absolute value of $\Phi_0$;
for $\Phi_0 = 0.0$ the effect is observed first and is
followed by $\Phi_0$-values of $\pm 0.1$ and $\pm 0.2$,
respectively.
For boson background fields $\Phi_0 \approx 1$ the system
becomes unstable only for relatively strong couplings
$g_R/m \approx 1.5$, which is in the regime of critical
coupling constants as observed in the 3-point approximation.
Thus the discussed effect in the 4-point approximation
is limited to small background fields where quantum
fluctuations are most prominent.
\section{Summary}
In the present work we have investigated the groundstate 
properties of the Yukawa-theory in $1+1$ dimensions 
using the method of correlation dynamics.
We derived the infinite coupled hierarchy of 
equations of motion for connected equal-time Green 
functions and introduced three different truncation
schemes, i.e. the 2-point-, 3-point- and 4-point approximation.
Furtheron, we discussed the necessary renormalization
prescription for the Yukawa-theory in $1+1$ dimensions
which -- besides an implicit counterterm renormalization
of the boson source -- consists in a multiplicative
renormalization of the coupling constant.

In our numerical calculations we found the following results:
\begin{itemize}
\item{
The generation of a correlated eigenstate of the interacting 
theory is possible by adiabatically switching-on the coupling
constant (in line with the Gell-Mann and Low theorem). }
\item{
The effective potentials in $Y_{1+1}CD(2)$-
and discretized GEP-approximation differ in the regime of
``negative effective fermion masses'' in case of a
finite basis set due to an infrared- and ultraviolet-cutoff.
These differences vanish in the infrared as well as in the
ultraviolet limit and are due to the noninclusion of
zero mode pair production in $Y_{1+1}CD(2)$-approximation.}
\item{In the $Y_{1+1}CD(3)$-approximation the propagation
breaks down in the non-convex region of the effective potential
where the continuum theory becomes unbounded from below; 
this is also known from the correlation dynamical
investigations of $\Phi^4$-theory and can be attributed to
tunneling \cite{phc97}.}
\item{In the $Y_{1+1}CD(4)$-approximation a novel effect
is observed, which has no analogon in the pure bosonic 
theories investigated so far.
Here the propagation breaks down already for small
couplings $g_R/m$, which is no longer a consequence of
the transition to the non-convex parameter regime, but
due to strong correlations induced by higher order connected 
Green functions in the fermionic sector.
The associated lowering of the single-particle energies
leads to a strong population of the previously 
weakly occupied zero momentum mode and finally to an
increasing instability of the system.
We speculate that this new effect is due to a condensation
of fermion-antifermion pairs, but cannot proove this explicitly 
due to the limited set of basis states that can be employed
numerically.}
\end{itemize}
The last effect, that persists in the ultraviolet limit,
shows that novel phenomena appear in a theory with
fermionic degrees of freedom since such a
behaviour was not observed in any pure boson theory
so far.
Furthermore, it demonstrates the importance of quantum 
correlations of higher order especially for low or
vanishing background fields $\Phi_0$.
It remains to be seen if such effects will persist also
in gauge field theories such as QCD where the correlation
dynamical approach has been formulated in \cite{wch95}.
\newpage
\begin{appendix}
\section{Cluster expansion of the mixed sector}
\label{clust}
%
%
%
%
%
%
%

%
%
\begin{eqnarray}
\lefteqn{
\langle \, \Psb(\vec{x}_1) \, \Psi(\vec{x}_2) \, 
      \hat{O}_3(\vec{x}_3) \, \hat{O}_4(\vec{x}_4) \, \rangle \; = \;
      \langle \, \Psb(\vec{x}_1) \, \Psi(\vec{x}_2) \,
              \hat{O}_3(\vec{x}_3) \, \hat{O}_4(\vec{x}_4) \, \rangle_c }
      \nonumber\\
& & + \langle \, \Psb(\vec{x}_1) \, \Psi(\vec{x}_2) \,
              \hat{O}_3(\vec{x}_3) \, \rangle_c \:
      \langle \, \hat{O}_4(\vec{x}_4) \, \rangle
      \nonumber\\
& & + \langle \, \Psb(\vec{x}_1) \, \Psi(\vec{x}_2) \, 
              \hat{O}_4(\vec{x}_4) \, \rangle_c \:
      \langle \, \hat{O}_3(\vec{x}_3) \, \rangle
      \nonumber\\
& & + \langle \, \Psb(\vec{x}_1) \, \Psi(\vec{x}_2) \, \rangle_c \:
      \langle \, \hat{O}_3(\vec{x}_3) \, \hat{O}_4(\vec{x}_4) \, \rangle_c
      \nonumber\\
& & + \langle \, \Psb(\vec{x}_1) \, \Psi(\vec{x}_2) \, \rangle_c \:
      \langle \, \hat{O}_3(\vec{x}_3) \, \rangle \:
      \langle \, \hat{O}_4(\vec{x}_4) \, \rangle
\label{twvffclust}
\end{eqnarray}
%
%

%
%
\begin{eqnarray}
\lefteqn{
      \langle \, \Psb(\vec{x}_1) \, \Psi(\vec{x}_2) \, 
            \hat{O}_3(\vec{x}_3) \, \hat{O}_4(\vec{x}_4) \,
            \hat{O}_5(\vec{x}_5) \, \rangle \; = \;
      \langle \, \Psb(\vec{x}_1) \, \Psi(\vec{x}_2) \,
            \hat{O}_3(\vec{x}_3) \, \hat{O}_4(\vec{x}_4) \,
            \hat{O}_5(\vec{x}_5) \, \rangle_c }
      \nonumber\\
& & + \langle \, \Psb(\vec{x}_1) \, \Psi(\vec{x}_2) \,
            \hat{O}_3(\vec{x}_3) \, \hat{O}_4(\vec{x}_4) \, \rangle_c \:
      \langle \, \hat{O}_5(\vec{x}_5) \, \rangle
      \nonumber\\
& & + \langle \, \Psb(\vec{x}_1) \, \Psi(\vec{x}_2) \, 
            \hat{O}_3(\vec{x}_3) \, \hat{O}_5(\vec{x}_5) \, \rangle_c \:
      \langle \, \hat{O}_4(\vec{x}_4) \, \rangle
      \nonumber\\
& & + \langle \, \Psb(\vec{x}_1) \, \Psi(\vec{x}_2) \, 
            \hat{O}_4(\vec{x}_4) \, \hat{O}_5(\vec{x}_5) \, \rangle_c \:
      \langle \, \hat{O}_3(\vec{x}_3) \, \rangle
      \nonumber\\
& & + \langle \, \Psb(\vec{x}_1) \, \Psi(\vec{x}_2) \, 
            \hat{O}_3(\vec{x}_3) \, \rangle_c \:
      \langle \, \hat{O}_4(\vec{x}_4) \, 
                 \hat{O}_5(\vec{x}_5) \, \rangle_c
      \nonumber\\
& & + \langle \, \Psb(\vec{x}_1) \, \Psi(\vec{x}_2) \,
            \hat{O}_3(\vec{x}_3) \, \rangle_c \:
      \langle \, \hat{O}_4(\vec{x}_4) \, \rangle \: 
      \langle \, \hat{O}_5(\vec{x}_5) \, \rangle
      \nonumber\\
& & + \langle \, \Psb(\vec{x}_1) \, \Psi(\vec{x}_2) \,
            \hat{O}_4(\vec{x}_4) \, \rangle_c \:
      \langle \, \hat{O}_3(\vec{x}_3) \,
                 \hat{O}_5(\vec{x}_5) \, \rangle_c
      \nonumber\\
& & + \langle \, \Psb(\vec{x}_1) \, \Psi(\vec{x}_2) \,
            \hat{O}_4(\vec{x}_4) \, \rangle_c \:
      \langle \, \hat{O}_3(\vec{x}_3) \, \rangle \:
      \langle \, \hat{O}_5(\vec{x}_5) \, \rangle
      \nonumber\\
& & + \langle \, \Psb(\vec{x}_1) \, \Psi(\vec{x}_2) \,
            \hat{O}_5(\vec{x}_5) \, \rangle_c \:
      \langle \, \hat{O}_3(\vec{x}_3) \, 
                 \hat{O}_4(\vec{x}_4) \, \rangle_c
      \nonumber\\
& & + \langle \, \Psb(\vec{x}_1) \, \Psi(\vec{x}_2) \, 
            \hat{O}_5(\vec{x}_5) \, \rangle_c \:
      \langle \, \hat{O}_3(\vec{x}_3) \, \rangle \:
      \langle \, \hat{O}_4(\vec{x}_4) \, \rangle
      \nonumber\\
& & + \langle \, \Psb(\vec{x}_1) \, \Psi(\vec{x}_2) \, \rangle_c \:
      \langle \, \hat{O}_3(\vec{x}_3) \, 
                 \hat{O}_4(\vec{x}_4) \,
                 \hat{O}_5(\vec{x}_5) \, \rangle_c
      \nonumber\\
& & + \langle \, \Psb(\vec{x}_1) \, \Psi(\vec{x}_2) \, \rangle_c \:
      \langle \, \hat{O}_3(\vec{x}_3) \, 
                 \hat{O}_4(\vec{x}_4) \, \rangle_c \:
      \langle \, \hat{O}_5(\vec{x}_5) \, \rangle
      \nonumber\\
& & + \langle \, \Psb(\vec{x}_1) \, \Psi(\vec{x}_2) \, \rangle_c \:
      \langle \, \hat{O}_3(\vec{x}_3) \,
                 \hat{O}_5(\vec{x}_5) \, \rangle_c \:
      \langle \, \hat{O}_4(\vec{x}_4) \, \rangle
      \nonumber\\
& & + \langle \, \Psb(\vec{x}_1) \, \Psi(\vec{x}_2) \, \rangle_c \:
      \langle \, \hat{O}_4(\vec{x}_4) \, 
                 \hat{O}_5(\vec{x}_5) \, \rangle_c \:
      \langle \, \hat{O}_3(\vec{x}_3) \, \rangle
      \nonumber\\
& & + \langle \, \Psb(\vec{x}_1) \, \Psi(\vec{x}_2) \, \rangle_c \:
      \langle \, \hat{O}_3(\vec{x}_3) \, \rangle \:
      \langle \, \hat{O}_4(\vec{x}_4) \, \rangle \:
      \langle \, \hat{O}_5(\vec{x}_5) \, \rangle
\label{twvfffclust}
\end{eqnarray}
%
%

%
%
\begin{eqnarray}
\lefteqn{
      \langle \, \Psb(\vec{x}_1) \, \Psb(\vec{x}_2) \,
                 \Psi(\vec{x}_3) \, \Psi(\vec{x}_4) \, 
            \hat{O}_5(\vec{x}_5) \, \rangle \; = \;
      \langle \, \Psb(\vec{x}_1) \, \Psb(\vec{x}_2) \,
                 \Psi(\vec{x}_3) \, \Psi(\vec{x}_4) \, 
            \hat{O}_5(\vec{x}_5) \, \rangle_c }
      \nonumber\\
& & + \langle \, \Psb(\vec{x}_1) \, \Psb(\vec{x}_2) \,
                 \Psi(\vec{x}_3) \, \Psi(\vec{x}_4) \, \rangle_c \:
      \langle \, \hat{O}_5(\vec{x}_5) \, \rangle
      \nonumber\\
& & + \langle \, \Psb(\vec{x}_1) \, \Psi(\vec{x}_4) \, 
            \hat{O}_5(\vec{x}_5) \, \rangle_c \:
      \langle \, \Psb(\vec{x}_2) \, \Psi(\vec{x}_3) \, \rangle_c \:
      \nonumber\\
& & + \langle \, \Psb(\vec{x}_1) \, \Psi(\vec{x}_4) \, \rangle_c \:
      \langle \, \Psb(\vec{x}_2) \, \Psi(\vec{x}_3) \,
            \hat{O}_5(\vec{x}_5) \, \rangle_c
      \nonumber\\
& & + \langle \, \Psb(\vec{x}_1) \, \Psi(\vec{x}_4) \, \rangle_c \:
      \langle \, \Psb(\vec{x}_2) \, \Psi(\vec{x}_3) \, \rangle_c \:
      \langle \, \hat{O}_5(\vec{x}_5) \, \rangle
      \nonumber\\
& & - \langle \, \Psb(\vec{x}_1) \, \Psi(\vec{x}_3) \, 
            \hat{O}_5(\vec{x}_5) \, \rangle_c \:
      \langle \, \Psb(\vec{x}_2) \, \Psi(\vec{x}_4) \, \rangle_c
      \nonumber\\
& & - \langle \, \Psb(\vec{x}_1) \, \Psi(\vec{x}_3) \, \rangle_c \:
      \langle \, \Psb(\vec{x}_2) \, \Psi(\vec{x}_4) \,
            \hat{O}_5(\vec{x}_5) \, \rangle_c
      \nonumber\\
& & - \langle \, \Psb(\vec{x}_1) \, \Psi(\vec{x}_3) \, \rangle_c \:
      \langle \, \Psb(\vec{x}_2) \, \Psi(\vec{x}_4) \, \rangle_c \:
      \langle \, \hat{O}_5(\vec{x}_5) \, \rangle
\label{twwvvfclust}
\end{eqnarray}
\section{Hierarchy of equations of motion for
connected equal-time Green functions in a
single-particle basis}
\label{appeom}
%
%
%
%
%
%
%
\begin{eqnarray}
\partial_t \, \langle \, \varphi_\alpha
             \, \varphi_\beta \, \varphi_\gamma \, \rangle_c
& = & \langle \, \pi_\alpha \, \varphi_\beta \, \varphi_\gamma \, \rangle_c
     + \:  \langle \, \varphi_\alpha \, \pi_\beta \, \varphi_\gamma \, \rangle_c
     + \:  \langle \, \varphi_\alpha \, \varphi_\beta \, \pi_\gamma \, \rangle_c
\label{opbfff}
\end{eqnarray}
\begin{eqnarray}
\partial_t \, \langle \, \pi_\alpha
           \, \varphi_\beta \, \varphi_\gamma \, \rangle_c
& = & \langle \, \pi_\alpha \, \pi_\beta \, \varphi_\gamma \, \rangle_c
     + \:  \langle \, \pi_\alpha \, \varphi_\beta \, \pi_\gamma \, \rangle_c
     + \:  \sum_{\lambda} \ t_{\alpha \lambda} \
      \langle \, \varphi_\lambda \, \varphi_\beta \, \varphi_\gamma \, \rangle_c
      \nonumber\\
& & - \: g_B \sum_{\lambda_1 \lambda_2} \
      \langle \alpha|\lambda_1\lambda_2 \rangle \
      \langle \, \psb_{i \lambda_1} \, \psi_{i \lambda_2}
      \, \varphi_\beta \, \varphi_\gamma \, \rangle_c
\label{opbpff}
\end{eqnarray}
\begin{eqnarray}
\partial_t \, \langle \, \pi_\alpha
             \, \pi_\beta \, \varphi_\gamma \, \rangle_c
& = & \langle \, \pi_\alpha \, \pi_\beta \, \pi_\gamma \, \rangle_c
     + \:  \sum_{\lambda} \ t_{\alpha \lambda} \
      \langle \, \varphi_\lambda \, \pi_\beta \, \varphi_\gamma \, \rangle_c \
     + \:  t_{\beta \lambda} \
      \langle \, \pi_\alpha \, \varphi_\lambda \, \varphi_\gamma \, \rangle_c
      \nonumber\\
& & - \: g_B \sum_{\lambda_1 \lambda_2} \
      \langle \alpha|\lambda_1\lambda_2 \rangle \
      \langle \, \psb_{i \lambda_1} \, \psi_{i \lambda_2}
      \, \pi_\beta \, \varphi_\gamma \, \rangle_c
      \nonumber\\
& & - \: g_B \sum_{\lambda_1 \lambda_2} \
      \langle \beta|\lambda_1\lambda_2 \rangle \
      \langle \, \psb_{i \lambda_1} \, \psi_{i \lambda_2}
      \, \pi_\alpha \, \varphi_\gamma \, \rangle_c
\label{opbppf}
\end{eqnarray}
\begin{eqnarray}
\partial_t \, \langle \, \pi_\alpha
             \, \pi_\beta \, \pi_\gamma \, \rangle_c
& = & \sum_{\lambda} \ t_{\alpha \lambda} \
      \langle \, \varphi_\lambda \, \pi_\beta \, \pi_\gamma \, \rangle_c \
     + \:  t_{\beta \lambda} \
      \langle \, \pi_\alpha \, \varphi_\lambda \, \pi_\gamma \, \rangle_c \
     + \:  t_{\gamma \lambda} \
      \langle \, \pi_\alpha \, \pi_\lambda \, \varphi_\gamma \, \rangle_c \
      \nonumber\\
& & - \: g_B \sum_{\lambda_1 \lambda_2} \
      \langle \alpha|\lambda_1\lambda_2 \rangle \
      \langle \, \psb_{i \lambda_1} \, \psi_{i \lambda_2}
      \, \pi_\beta \, \pi_\gamma \, \rangle_c
      \nonumber\\
& & - \: g_B \sum_{\lambda_1 \lambda_2} \
      \langle \beta|\lambda_1\lambda_2 \rangle \
      \langle \, \psb_{i \lambda_1} \, \psi_{i \lambda_2}
      \, \pi_\alpha \, \pi_\gamma \, \rangle_c
      \nonumber\\
& & - \: g_B \sum_{\lambda_1 \lambda_2} \
      \langle \gamma|\lambda_1\lambda_2 \rangle \
      \langle \, \psb_{i \lambda_1} \, \psi_{i \lambda_2}
      \, \pi_\alpha \, \pi_\beta \, \rangle_c
\label{opbppp}
\end{eqnarray}
%
%
%
%
\begin{eqnarray}
\partial_t \, \langle \, \psb_{a\alpha} \, \psi_{b\beta} \, \varphi_\gamma \, \rangle_c
& = & \sum_{\lambda} \ \tilde{t}_{\lambda\alpha}^{la} \
      \langle \, \psb_{l\lambda} \, \psi_{b\beta} \, \varphi_\gamma \, \rangle_c \
    - \: \tilde{t}_{\beta\lambda}^{bl} \
      \langle \, \psb_{a\alpha} \, \psi_{l \lambda} \, \varphi_\gamma \, \rangle_c
      \nonumber\\
& &  + \:  i g_B \beta_{la} \sum_{\lambda_1 \lambda_2} \
      \langle \alpha|\lambda_1\lambda_2 \rangle \
      \langle \, \psb_{l \lambda_1} \, \psi_{b \beta}
              \, \varphi_{\lambda_2} \, \varphi_\gamma \, \rangle_c
      \nonumber\\
& & - \: i g_B \beta_{bl} \sum_{\lambda_1 \lambda_2} \
      \langle \beta|\lambda_1\lambda_2 \rangle \
      \langle \, \psb_{a \alpha} \, \psi_{l \lambda_1}
              \, \varphi_{\lambda_2} \, \varphi_\gamma \, \rangle_c
      \nonumber\\
& &  + \:  i g_B \beta_{la} \sum_{\lambda_1 \lambda_2} \
      \langle \alpha|\lambda_1\lambda_2 \rangle \
      \langle \, \psb_{l \lambda_1} \, \psi_{b \beta} \, \rangle_c \
      \langle \, \varphi_{\lambda_2} \, \varphi_\gamma \, \rangle_c
      \nonumber\\
& & - \: i g_B \beta_{bl} \sum_{\lambda_1 \lambda_2} \
      \langle \beta|\lambda_1\lambda_2 \rangle \
      \langle \, \psb_{a \alpha} \, \psi_{l \lambda_1} \, \rangle_c \
      \langle \, \varphi_{\lambda_2} \, \varphi_\gamma \, \rangle_c
      \nonumber\\
& &  + \:  \langle \, \psb_{a \alpha} \, \psi_{b \beta}
              \, \pi_\gamma \, \rangle_c
\label{opbwvf}
\end{eqnarray}
\begin{eqnarray}
\partial_t \, \langle \, \psb_{a\alpha} \, \psi_{b\beta} \, \pi_\gamma \, \rangle_c
& = & \phantom{- \: } \sum_{\lambda} \ \tilde{t}_{\lambda\alpha}^{la} \
      \langle \, \psb_{l\lambda} \, \psi_{b\beta} \, \pi_\gamma \, \rangle_c \
    - \: \tilde{t}_{\beta\lambda}^{bl} \
      \langle \, \psb_{a\alpha} \, \psi_{l \lambda} \, \pi_\gamma \, \rangle_c
      \nonumber\\
& &  + \:  \sum_{\lambda} \ t_{\gamma \lambda} \
      \langle \, \psb_{a\alpha} \, \psi_{b \beta} \, \varphi_\lambda \, \rangle_c
      \nonumber\\
& &  + \:  i g_B \beta_{la} \sum_{\lambda_1 \lambda_2} \
      \langle \alpha|\lambda_1\lambda_2 \rangle \
      \langle \, \psb_{l \lambda_1} \, \psi_{b \beta}
              \, \varphi_{\lambda_2} \, \pi_\gamma \, \rangle_c
      \nonumber\\
& & - \: i g_B \beta_{bl} \sum_{\lambda_1 \lambda_2} \
      \langle \beta|\lambda_1\lambda_2 \rangle \
      \langle \, \psb_{a \alpha} \, \psi_{l \lambda_1}
              \, \varphi_{\lambda_2} \, \pi_\gamma \, \rangle_c
      \nonumber\\
& &  + \:  i g_B \beta_{la} \sum_{\lambda_1 \lambda_2} \
      \langle \alpha|\lambda_1\lambda_2 \rangle \
      \langle \, \psb_{l \lambda_1} \, \psi_{b \beta} \, \rangle_c \
      \langle \, \varphi_{\lambda_2} \, \pi_\gamma \, \rangle_c
      \nonumber\\
& & - \: i g_B \beta_{bl} \sum_{\lambda_1 \lambda_2} \
      \langle \beta|\lambda_1\lambda_2 \rangle \
      \langle \, \psb_{a \alpha} \, \psi_{l \lambda_1} \, \rangle_c \
      \langle \, \varphi_{\lambda_2} \, \pi_\gamma \, \rangle_c
      \nonumber\\
& & - \: g_B \sum_{\lambda_1 \lambda_2} \
      \langle \gamma|\lambda_1\lambda_2 \rangle \
      \langle \, \psb_{a \alpha} \, \psi_{b \beta}
              \, \psb_{i \lambda_1} \, \psi_{i \lambda_2} \, \rangle_c
      \nonumber\\
& & - \: g_B \sum_{\lambda_1 \lambda_2} \
      \langle \gamma|\lambda_1\lambda_2 \rangle \
      \langle \, \psb_{a \alpha} \, \psi_{i \lambda_2} \, \rangle_c \
      \langle \, \psb_{i \lambda_1} \, \psi_{b \beta} \, \rangle_c
\label{opbwvp}
\end{eqnarray}
%
%
%
%
\begin{eqnarray}
\partial_t \, \langle \, \psb_{a\alpha} \, \psi_{b\beta}
                   \, \varphi_\gamma \, \varphi_\delta \, \rangle_c
& = & \sum_{\lambda} \ \tilde{t}_{\lambda\alpha}^{la} \
      \langle \, \psb_{l\lambda} \, \psi_{b\beta}
              \, \varphi_\gamma \, \varphi_\delta \, \rangle_c \
    - \: \tilde{t}_{\beta\lambda}^{bl}
      \langle \, \psb_{a\alpha} \, \psi_{l \lambda}
              \, \varphi_\gamma \, \varphi_\delta \, \rangle_c
      \nonumber\\
& &  + \:  i g_B \beta_{la} \sum_{\lambda_1 \lambda_2} \
      \langle \alpha|\lambda_1\lambda_2 \rangle \
      \langle \, \psb_{l \lambda_1} \, \psi_{b \beta}
              \, \varphi_{\lambda_2} \, \varphi_\gamma \, \varphi_\delta \, \rangle_c
      \nonumber\\
& &  + \:  i g_B \beta_{la} \sum_{\lambda_1 \lambda_2} \
      \langle \alpha|\lambda_1\lambda_2 \rangle \
      \langle \, \psb_{l \lambda_1} \, \psi_{b \beta} \, \varphi_\gamma \, \rangle_c \
      \langle \, \varphi_{\lambda_2} \, \varphi_\delta \, \rangle_c
      \nonumber\\
& &  + \:  i g_B \beta_{la} \sum_{\lambda_1 \lambda_2} \
      \langle \alpha|\lambda_1\lambda_2 \rangle \
      \langle \, \psb_{l \lambda_1} \, \psi_{b \beta} \, \varphi_\delta \, \rangle_c \
      \langle \, \varphi_{\lambda_2} \, \varphi_\gamma \, \rangle_c
      \nonumber\\
& & - \: i g_B \beta_{bl} \sum_{\lambda_1 \lambda_2} \
      \langle \beta|\lambda_1\lambda_2 \rangle \
      \langle \, \psb_{a \alpha} \, \psi_{l \lambda_1}
              \, \varphi_{\lambda_2} \, \varphi_\gamma \, \varphi_\delta \, \rangle_c
      \nonumber\\
& & - \: i g_B \beta_{bl} \sum_{\lambda_1 \lambda_2} \
      \langle \beta|\lambda_1\lambda_2 \rangle \
      \langle \, \psb_{a \alpha} \, \psi_{l \lambda_1} \, \varphi_\gamma \, \rangle_c \
      \langle \, \varphi_{\lambda_2} \, \varphi_\delta \, \rangle_c
      \nonumber\\
& & - \: i g_B \beta_{bl} \sum_{\lambda_1 \lambda_2} \
      \langle \beta|\lambda_1\lambda_2 \rangle \
      \langle \, \psb_{a \alpha} \, \psi_{l \lambda_1} \, \varphi_\delta \, \rangle_c \
      \langle \, \varphi_{\lambda_2} \, \varphi_\gamma \, \rangle_c
      \nonumber\\
& &  + \:  \langle \, \psb_{a \alpha} \, \psi_{b \beta}
              \, \pi_\gamma \, \varphi_\delta \, \rangle_c
     + \:  \langle \, \psb_{a \alpha} \, \psi_{b \beta}
              \, \varphi_\gamma \, \pi_\delta \, \rangle_c
\label{opbwff}
\end{eqnarray}
\begin{eqnarray}
\partial_t \, \langle \, \psb_{a\alpha} \, \psi_{b\beta}
                   \, \pi_\gamma \, \varphi_\delta \, \rangle_c
& = & \sum_{\lambda} \ \tilde{t}_{\lambda\alpha}^{la} \
      \langle \, \psb_{l\lambda} \, \psi_{b\beta}
              \, \pi_\gamma \, \varphi_\delta \, \rangle_c \
    - \: \tilde{t}_{\beta\lambda}^{bl} \
      \langle \, \psb_{a\alpha} \, \psi_{l \lambda}
              \, \pi_\gamma \, \varphi_\delta \, \rangle_c
      \nonumber\\
& &  + \:  \langle \, \psb_{a\alpha} \, \psi_{b \beta}
              \, \pi_\lambda \, \pi_\delta \, \rangle_c
     + \:  \sum_{\lambda} \ t_{\gamma \lambda} \
      \langle \, \psb_{a\alpha} \, \psi_{b \beta}
              \, \varphi_\lambda \, \varphi_\delta \, \rangle_c
      \nonumber\\
& &  + \:  i g_B \beta_{la} \sum_{\lambda_1 \lambda_2} \
      \langle \alpha|\lambda_1\lambda_2 \rangle \
      \langle \, \psb_{l \lambda_1} \, \psi_{b \beta}
              \, \varphi_{\lambda_2} \, \pi_\gamma \, \varphi_\delta \, \rangle_c
      \nonumber\\
& &  + \:  i g_B \beta_{la} \sum_{\lambda_1 \lambda_2} \
      \langle \alpha|\lambda_1\lambda_2 \rangle \
      \langle \, \psb_{l \lambda_1} \, \psi_{b \beta} \, \pi_\gamma \, \rangle_c \
      \langle \, \varphi_{\lambda_2} \, \varphi_\delta \, \rangle_c
      \nonumber\\
& &  + \:  i g_B \beta_{la} \sum_{\lambda_1 \lambda_2} \
      \langle \alpha|\lambda_1\lambda_2 \rangle \
      \langle \, \psb_{l \lambda_1} \, \psi_{b \beta} \, \varphi_\delta \, \rangle_c \
      \langle \, \varphi_{\lambda_2} \, \pi_\gamma \, \rangle_c
      \nonumber\\
& & - \: i g_B \beta_{bl} \sum_{\lambda_1 \lambda_2} \
      \langle \beta|\lambda_1\lambda_2 \rangle \
      \langle \, \psb_{a \alpha} \, \psi_{l \lambda_1}
              \, \varphi_{\lambda_2} \, \pi_\gamma \, \varphi_\delta \, \rangle_c
      \nonumber\\
& & - \: i g_B \beta_{bl} \sum_{\lambda_1 \lambda_2} \
      \langle \beta|\lambda_1\lambda_2 \rangle \
      \langle \, \psb_{a \alpha} \, \psi_{l \lambda_1} \, \pi_\gamma \, \rangle_c \
      \langle \, \varphi_{\lambda_2} \, \varphi_\delta \, \rangle_c
      \nonumber\\
& & - \: i g_B \beta_{bl} \sum_{\lambda_1 \lambda_2} \
      \langle \beta|\lambda_1\lambda_2 \rangle \
      \langle \, \psb_{a \alpha} \, \psi_{l \lambda_1} \, \varphi_\delta \, \rangle_c \
      \langle \, \varphi_{\lambda_2} \, \pi_\gamma \, \rangle_c
      \nonumber\\
& & - \: g_B \sum_{\lambda_1 \lambda_2} \
      \langle \gamma|\lambda_1\lambda_2 \rangle \
      \langle \, \psb_{a \alpha} \, \psi_{i \lambda_1} \, \varphi_\delta \, \rangle_c \
      \langle \, \psi_{b \beta} \, \psb_{i \lambda_2} \, \rangle_c
      \nonumber\\
& & - \: g_B \sum_{\lambda_1 \lambda_2} \
      \langle \gamma|\lambda_1\lambda_2 \rangle \
      \langle \, \psb_{a \alpha} \, \psi_{i \lambda_1} \, \rangle_c \
      \langle \, \psi_{b \beta} \, \psb_{i \lambda_2} \, \pi_\delta \, \rangle_c
\label{opbwvpf}
\end{eqnarray}
\begin{eqnarray}
\partial_t \, \langle \, \psb_{a\alpha} \, \psi_{b\beta}
                   \, \pi_\gamma \, \pi_\delta \, \rangle_c
& = & \phantom{+ \: } \sum_{\lambda} \ \tilde{t}_{\lambda\alpha}^{la} \
      \langle \, \psb_{l\lambda} \, \psi_{b\beta}
              \, \pi_\gamma \, \pi_\delta \, \rangle_c \
    - \: \tilde{t}_{\beta\lambda}^{bl} \
      \langle \, \psb_{a\alpha} \, \psi_{l \lambda} \
              \, \pi_\gamma \, \pi_\delta \, \rangle_c
      \nonumber\\
& &  + \:  \sum_{\lambda} \ t_{\gamma \lambda} \
      \langle \, \psb_{a\alpha} \, \psi_{b \beta}
              \, \varphi_\lambda \, \pi_\delta \, \rangle_c \
     + \:  t_{\delta \lambda} \
      \langle \, \psb_{a\alpha} \, \psi_{b \beta}
              \, \pi_\gamma \, \varphi_\lambda \, \rangle_c
      \nonumber\\
& &  + \:  i g_B \beta_{la} \sum_{\lambda_1 \lambda_2} \
      \langle \alpha|\lambda_1\lambda_2 \rangle \
      \langle \, \psb_{l \lambda_1} \, \psi_{b \beta}
              \, \varphi_{\lambda_2} \, \pi_\gamma \, \pi_\delta \, \rangle_c
      \nonumber\\
& &  + \:  i g_B \beta_{la} \sum_{\lambda_1 \lambda_2} \
      \langle \alpha|\lambda_1\lambda_2 \rangle \
      \langle \, \psb_{l \lambda_1} \, \psi_{b \beta} \, \pi_\gamma \, \rangle_c \
      \langle \, \varphi_{\lambda_2} \, \pi_\delta \, \rangle_c
      \nonumber\\
& &  + \:  i g_B \beta_{la} \sum_{\lambda_1 \lambda_2} \
      \langle \alpha|\lambda_1\lambda_2 \rangle \
      \langle \, \psb_{l \lambda_1} \, \psi_{b \beta} \, \pi_\delta \, \rangle_c \
      \langle \, \varphi_{\lambda_2} \, \pi_\gamma \, \rangle_c
      \nonumber\\
& & - \: i g_B \beta_{bl} \sum_{\lambda_1 \lambda_2} \
      \langle \beta|\lambda_1\lambda_2 \rangle \
      \langle \, \psb_{a \alpha} \, \psi_{l \lambda_1}
              \, \varphi_{\lambda_2} \, \pi_\gamma \, \pi_\delta \, \rangle_c
      \nonumber\\
& & - \: i g_B \beta_{bl} \sum_{\lambda_1 \lambda_2} \
      \langle \beta|\lambda_1\lambda_2 \rangle \
      \langle \, \psb_{a \alpha} \, \psi_{l \lambda_1} \, \pi_\gamma \, \rangle_c \
      \langle \, \varphi_{\lambda_2} \, \pi_\delta \, \rangle_c
      \nonumber\\
& & - \: i g_B \beta_{bl} \sum_{\lambda_1 \lambda_2} \
      \langle \beta|\lambda_1\lambda_2 \rangle \
      \langle \, \psb_{a \alpha} \, \psi_{l \lambda_1} \, \pi_\delta \, \rangle_c \
      \langle \, \varphi_{\lambda_2} \, \pi_\gamma \, \rangle_c
      \nonumber\\
& & - \: g_B \sum_{\lambda_1 \lambda_2} \
      \langle \gamma|\lambda_1\lambda_2 \rangle \
      \langle \, \psb_{a \alpha} \, \psi_{i \lambda_1} \, \pi_\delta \, \rangle_c \
      \langle \, \psi_{b \beta} \, \psb_{i \lambda_2} \, \rangle_c
      \nonumber\\
& & - \: g_B \sum_{\lambda_1 \lambda_2} \
      \langle \gamma|\lambda_1\lambda_2 \rangle \
      \langle \, \psb_{a \alpha} \, \psi_{i \lambda_1} \, \rangle_c \
      \langle \, \psi_{b \beta} \, \psb_{i \lambda_2} \, \pi_\delta \, \rangle_c
      \nonumber\\
& & - \: g_B \sum_{\lambda_1 \lambda_2} \
      \langle \delta|\lambda_1\lambda_2 \rangle \
      \langle \, \psb_{a \alpha} \, \psi_{i \lambda_1} \, \pi_\gamma \, \rangle_c \
      \langle \, \psi_{b \beta} \, \psb_{i \lambda_2} \, \rangle_c
      \nonumber\\
& & - \: g_B \sum_{\lambda_1 \lambda_2} \
      \langle \delta|\lambda_1\lambda_2 \rangle \
      \langle \, \psb_{a \alpha} \, \psi_{i \lambda_1} \, \rangle_c \
      \langle \, \psi_{b \beta} \, \psb_{i \lambda_2} \, \pi_\gamma \, \rangle_c
\label{opbwvpp}
\end{eqnarray}
\begin{eqnarray}
\partial_t \, \langle \, \psb_{a\alpha} \, \psb_{b\beta}
                   \, \psi_{c\gamma} \, \psi_{d\delta} \, \rangle_c
& = & \phantom{ - \: } \sum_{\lambda} \ \tilde{t}_{\lambda\alpha}^{la} \
      \langle \, \psb_{l\lambda} \, \psb_{b\beta}
              \, \psi_{c \gamma} \, \psi_{d \delta} \, \rangle_c \
     + \:  \tilde{t}_{\lambda\alpha}^{lb} \
      \langle \, \psb_{a \alpha} \, \psb_{b \lambda}
              \, \psi_{c \gamma} \, \psi_{d \delta} \, \rangle_c
      \nonumber\\
& & - \: \sum_{\lambda} \ \tilde{t}_{\gamma\lambda}^{cl} \
      \langle \, \psb_{a \alpha} \, \psb_{b \beta}
              \, \psi_{l \lambda} \, \psi_{d \delta} \, \rangle_c \
     + \:  \tilde{t}_{\delta\lambda}^{dl} \
      \langle \, \psb_{a \alpha} \, \psb_{b \beta}
              \, \psi_{c \gamma} \, \psi_{l \lambda} \, \rangle_c
      \nonumber\\
& &  + \:  i g_B \beta_{la} \sum_{\lambda_1 \lambda_2} \
      \langle \alpha|\lambda_1\lambda_2 \rangle \
      \langle \, \psb_{l \lambda_1} \, \psi_{d \delta} \, \rangle_c \
      \langle \, \psb_{b \beta} \, \psi_{c \gamma} \, \varphi_{\lambda_2} \, \rangle_c
      \nonumber\\
& & - \: i g_B \beta_{la} \sum_{\lambda_1 \lambda_2} \
      \langle \alpha|\lambda_1\lambda_2 \rangle \
      \langle \, \psb_{l \lambda_1} \, \psi_{c \gamma} \, \rangle_c \
      \langle \, \psb_{b \beta} \, \psi_{d \delta} \, \varphi_{\lambda_2} \, \rangle_c
      \nonumber\\
& &  + \:  i g_B \beta_{lb} \sum_{\lambda_1 \lambda_2} \
      \langle \beta|\lambda_1\lambda_2 \rangle \
      \langle \, \psb_{l \lambda_1} \, \psi_{c \gamma} \, \rangle_c \
      \langle \, \psb_{a \alpha} \, \psi_{d \delta} \, \varphi_{\lambda_2} \, \rangle_c
      \nonumber\\
& & - \: i g_B \beta_{lb} \sum_{\lambda_1 \lambda_2} \
      \langle \beta|\lambda_1\lambda_2 \rangle \
      \langle \, \psb_{l \lambda_1} \, \psi_{d \delta} \, \rangle_c \
      \langle \, \psb_{a \alpha} \, \psi_{c \gamma} \, \varphi_{\lambda_2} \, \rangle_c
      \nonumber\\
& &  + \:  i g_B \beta_{cl} \sum_{\lambda_1 \lambda_2} \
      \langle \gamma|\lambda_1\lambda_2 \rangle \
      \langle \, \psb_{a \alpha} \, \psi_{l \lambda_1} \, \rangle_c \
      \langle \, \psb_{b \beta} \, \psi_{d \delta} \, \varphi_{\lambda_2} \, \rangle_c
      \nonumber\\
& & - \: i g_B \beta_{cl} \sum_{\lambda_1 \lambda_2} \
      \langle \gamma|\lambda_1\lambda_2 \rangle \
      \langle \, \psb_{b \beta} \, \psi_{l \lambda_1} \, \rangle_c \
      \langle \, \psb_{a \alpha} \, \psi_{d \delta} \, \varphi_{\lambda_2} \, \rangle_c
      \nonumber\\
& &  + \:  i g_B \beta_{dl} \sum_{\lambda_1 \lambda_2} \
      \langle \delta|\lambda_1\lambda_2 \rangle \
      \langle \, \psb_{b \beta} \, \psi_{l \lambda_1} \, \rangle_c \
      \langle \, \psb_{a \alpha} \, \psi_{c \gamma} \, \varphi_{\lambda_2} \, \rangle_c
      \nonumber\\
& & - \: i g_B \beta_{dl} \sum_{\lambda_1 \lambda_2} \
      \langle \delta|\lambda_1\lambda_2 \rangle \
      \langle \, \psb_{a \alpha} \, \psi_{l \lambda_1} \, \rangle_c \
     \langle \, \psb_{b \beta} \, \psi_{c \gamma} \, \varphi_{\lambda_2} \, \rangle_c
\label{opbwwvv}
\end{eqnarray}
\newpage
\section{Diagrammatical representation of the equation 
of motion for a connected equal-time 3-point function}

\begin{figure}[t]
{\vspace*{-2.0 cm}}
{\psfig{file=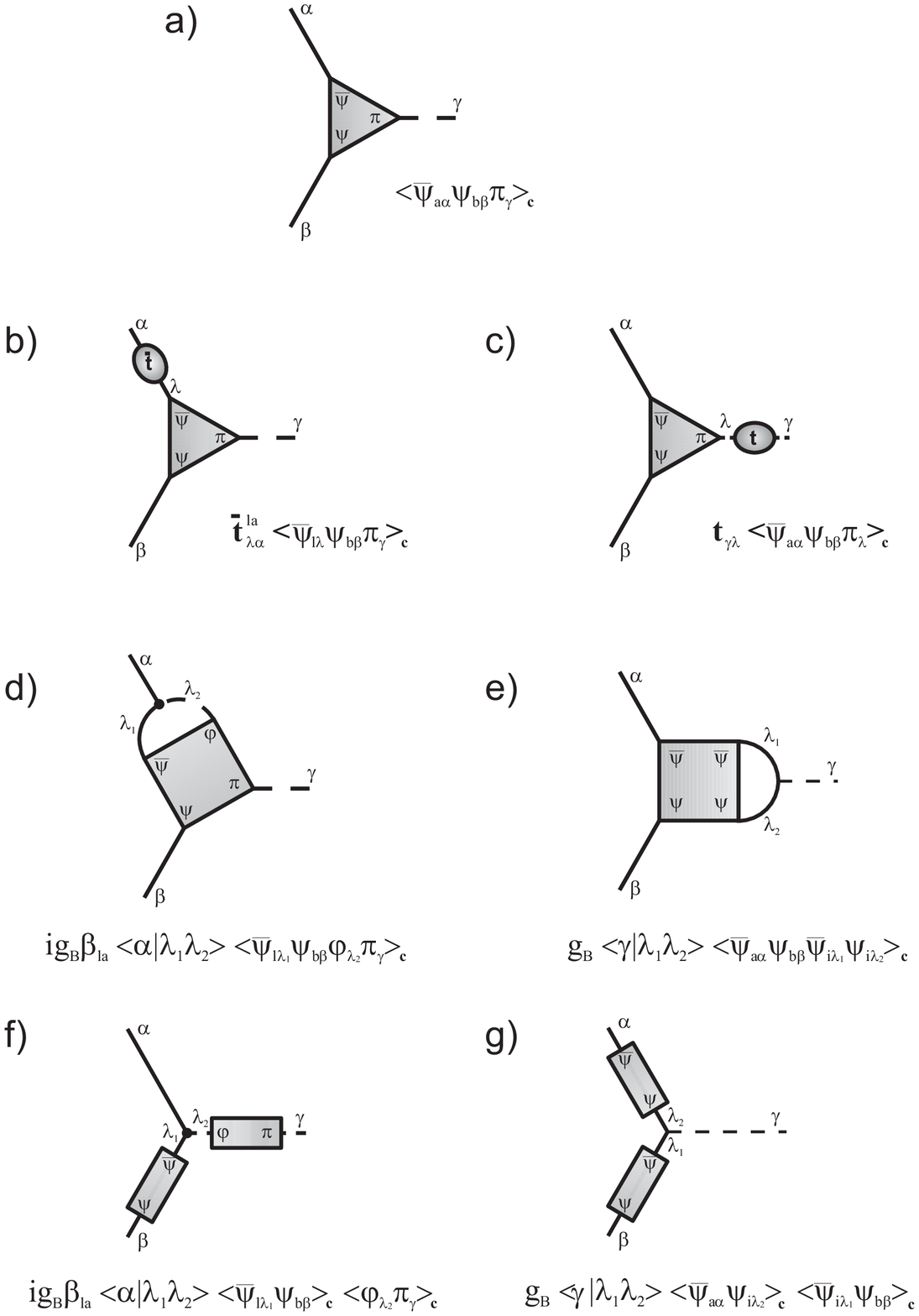,height=20.0 cm,width=15.0 cm}}
\caption{
\label{bild12}}
\end{figure}

$\phantom{in}$
To clearify the structure of the correlation dynamical
equations of motion we introduce a diagrammatical
representation. 
In Fig.~\ref{bild12} we demonstrate this for the equation of 
motion of the connected equal-time 3-point function 
$\langle \, \psb_{a\alpha} \, \psi_{b\beta} 
                           \, \pi_\gamma \, \rangle_c$,
where all contributions are displayed which arise 
from the $\Psb$- or the $\Pi$-part of the Green function 
(the contributions from field operator $\Psi$ have the same 
structure as the ones from $\Psb$ only modified by a relative 
minus-sign). 
The n-point connected equal-time Greenfunctions are represented by 
triangles or rectangles with a corresponding number of field 
operators while the greek letters at the lines --
fermionic lines: solid, bosonic lines: dashed --
going out of the connected Greenfunctions denote the momentum 
indices with respect to the chosen single-particle 
basis -- spinor indices are neglected for transparency.
We mention that all diagrams that emerge due to the application
of the cluster expansion on the equation of motion for the 
corresponding unconnected 3-point function contribute which are 
connected in the usual (Feynman-)sense, while all disconnected
diagrams cancel out when going over from the hierarchy
for unconnected Green functions to the hierarchy for connected 
Green functions.
\end{appendix}
\newpage
\newpage
\newcounter{figno}
{\usecounter{figno}\setlength{\rightmargin}{\leftmargin}}
\newpage
\newpage
\begin{figure}[hbtp]
{\vspace*{2.0 cm}}
{\psfig{file=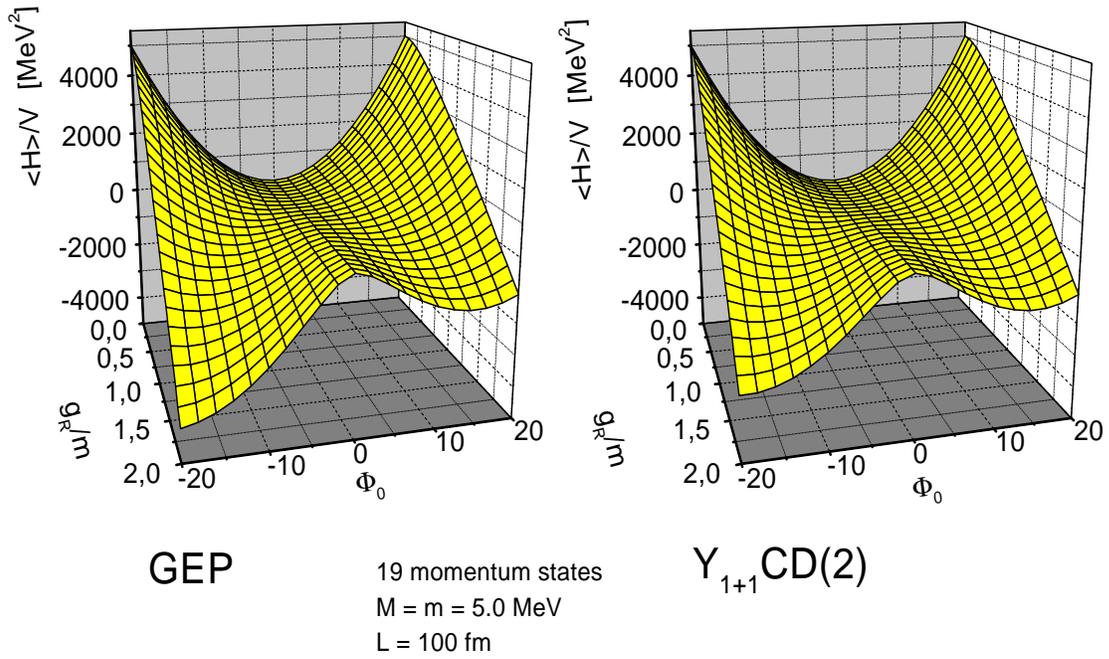,height=12.0 cm,width=16.0 cm}}
\caption{
Effective potential in $Y_{1+1}CD(2)$-approximation
as a function of $g_R/m$ and $\Phi_0$ (right part) and the
effective potential in the discretized GEP-approximation
(left part).
\label{bild3}}
\end{figure}
\newpage
\newpage
\newpage
\newpage
\newpage
\newpage
\newpage
\newpage
\newpage
\newpage
\begin{figure}[t]
{\center{ 
Figure 12 ;
Diagrammatical representation of the equation of motion
for the connected equal-time 3-point function
$\langle \, \psb_{a\alpha} \, \psi_{b\beta} 
                           \, \pi_\gamma \, \rangle_c$.
Connected equal-time n-point functions are represented by
triangles or rectangles with a corresponding number of
field-operators. The greek letters at the lines going
out from the connected Green functions  -- fermionic lines:
solid , bosonic lines: dashed --
denote the momentum indices with respect to the
chosen single particle basis (spinor indices are neglected
for clearness).
At the ``vertices'' momentum conservation is obeyed
according to (\ref{opbcontr}).
Displayed are all contributions to the time evolution
of $\langle \, \psb_{a\alpha} \, \psi_{b\beta}
                           \, \pi_\gamma \, \rangle_c$
(part a)) that arise from its $\psb$- (left diagrams) 
and $\pi$-terms (right diagrams) (cf. (\ref{opbwvp})).
Parts b), c) represent contributions stemming
from the free part of the Hamiltonian;
the t-symbols correspond to the operators
defined in (\ref{opbkinboson}) and (\ref{opbkinfermion});
Parts d), e) display the coupling to connected Green 
functions of order n+1 introduced by the Yukawa 
interaction;
Parts f), g) show the coupling to products of connected 
Green functions of lower order as a result of the cluster
expansion.}}
\end{figure}

\end{document}